\documentclass[12pt]{article}  
\usepackage{color}
\usepackage{ulem}
\usepackage{epsfig,amsfonts,amssymb}

\usepackage[english]{babel}
\usepackage{amsmath,amssymb}
\usepackage{amsfonts}
\usepackage{mathrsfs}
\usepackage{pgf,pgfarrows}
\usepackage{epic}
\usepackage{eepic}
\usepackage{color}
\usepackage[all,cmtip,line]{xy}
\usepackage{graphicx}
\usepackage{hyperref}

\newcommand{\comma}{\;\; ,}
\newcommand{\period}{\;\; .}
\newcommand{\eq}{\; = \;}
\newcommand{\sep}{\;\; , \;\;}

\newcommand{\be}{\begin{equation}}
\newcommand{\bd}{\begin{displaymath}}
\newcommand{\ee}{\end{equation}}
\newcommand{\ed}{\end{displaymath}}
\newcommand{\ba}{\begin{eqnarray}}
\newcommand{\ea}{\end{eqnarray}}

\renewcommand{\i}{{\rm i}}
\newcommand{\e}{{\rm e}}

\newcommand{\sn}{{\rm {sn}}}
\newcommand{\cn}{{\rm {cn}}}
\newcommand{\dn}{{\rm {dn}}}
\newcommand{\dv}{{\rm {d}}}
\newcommand{\bl}{\color{\ctnclr} }
\newcommand{\ccite}{\bl \cite}

\newcommand{\iqq}{{\rm i }}

\renewcommand{\i}{{\rm i}}

\renewcommand{\theequation}{\arabic{equation}}
\renewcommand{\theequation}{\arabic{equation}}
\newcommand{\ctnclr}{blue}
\newcommand{\om}{\omega}
\newcommand{\gom}{g(\e^{{\rm i}  \omega}) }
\renewcommand{\hom}{h(\e^{- {\rm i}  \omega}) }
\newcommand{\half}{\textstyle \frac{1}{2}}

\title{Onsager and Kaufman's calculation of the spontaneous 
magnetization of the Ising model }

\author{ R.J. Baxter\\
{\protect \small  Mathematical
Sciences Institute}\\
{\protect \small  The Australian National University,
 Canberra, A.C.T. 0200, Australia}\\
 }

\date{}

\begin{document}


\maketitle

\definecolor{magenta}{rgb}{0.5,0,0.5}

\definecolor{red}{rgb}{0.8,0,0}

\definecolor{green}{rgb}{0,0.5,0}

\definecolor{blue}{rgb}{0,0,0.8}

\vspace{2cm}

\definecolor{black}{rgb}{0.4,0.4,0.4}

\definecolor{brown}{rgb}{0.4,0.2,0.2}

 \setlength{\unitlength}{1pt}

 \abstract{Lars Onsager announced in 1949 that he and Bruria Kaufman
 had proved a simple formula for the  spontaneous 
magnetization of the square-lattice Ising model, but did not publish
their derivation. It was three years later when C. N. Yang published a 
derivation in Physical Review. In 1971 Onsager gave some clues to
his and Kaufman's method, and there are copies of their correspondence 
in 1950 now available on the Web and elsewhere. Here we review  how 
the calculation appears to have developed, and add a copy of a draft 
paper, almost certainly by  Onsager and Kaufman, that obtains the result.}

 \vspace{5mm}



 \section{Introduction}
 


\setcounter{equation}{0}
\renewcommand{\theequation}{\arabic{section}.\arabic{equation}}

Onsager calculated the free energy of the two-dimensional square-lattice
Ising model in 1944.{\ccite{Onsager1944}}  He did this by showing that the
transfer matrix is a product of two matrices, which together
generate (by successive commutations) a finite-dimensional Lie algebra
(the ``Onsager algebra''). In 1949  Bruria Kaufman 
gave a simpler 
derivation{\ccite{Kaufman1949}}   of this result, 
using anti-commuting  spinor (free-fermion) operators, i.e. a 
Clifford algebra.

Onsager was the Josiah Willard Gibbs Professor of 
Theoretical Chemistry at Yale University.
Kaufman had recently completed her PhD at Columbia University 
in New York, and was a research associate at the Institute
 for Advanced Study in Princeton.

\begin{figure}[hbt]
\begin{picture}(320,240) (-14,-40)

\setlength{\unitlength}{1.0pt}

{\color{black}

\put (60,10) {\line(1,0) {250}}
\put (60,11) {\line(1,0) {250}}

\put (60,60) {\line(1,0) {250}}
\put (60,61) {\line(1,0) {250}}

\put (60,110) {\line(1,0) {250}}
\put (60,111) {\line(1,0) {250}}

\put (60,160) {\line(1,0) {250}}
\put (60,161) {\line(1,0) {250}}

\put (85,-15) {\line(0,1) {200}}
\put (86,-15) {\line(0,1) {200}}

\put (135,-15) {\line(0,1) {200}}
\put (136,-15) {\line(0,1) {200}}

\put (185,-15) {\line(0,1) {200}}
\put (186,-15) {\line(0,1) {200}}

\put (235,-15) {\line(0,1) {200}}
\put (236,-15) {\line(0,1) {200}}

\put (285,-15) {\line(0,1) {200}}
\put (286,-15) {\line(0,1) {200}}

\multiput(85,10)(50,0){5}{\circle*{10}}
\multiput(85,60)(50,0){5}{\circle*{10}}
\multiput(85,110)(50,0){5}{\circle*{10}}
\multiput(85,160)(50,0){5}{\circle*{10}}

\put (85,-35){\large{1}}
\put (135,-35){\large{2}}
\put (235,-35){\Large{$i$}}

\put (320,5){\large{1}}
\put (320,55){\large{2}}
\put (320,105){\Large{$j$}}

\put (52,43){\large{(1,2)}}

\put (204,93){\large{$(i,j)$}}

}
\end{picture}

 \caption{\small The square lattice.} 
  \label{sqlatt}
  \vspace{1cm}

 \end{figure}
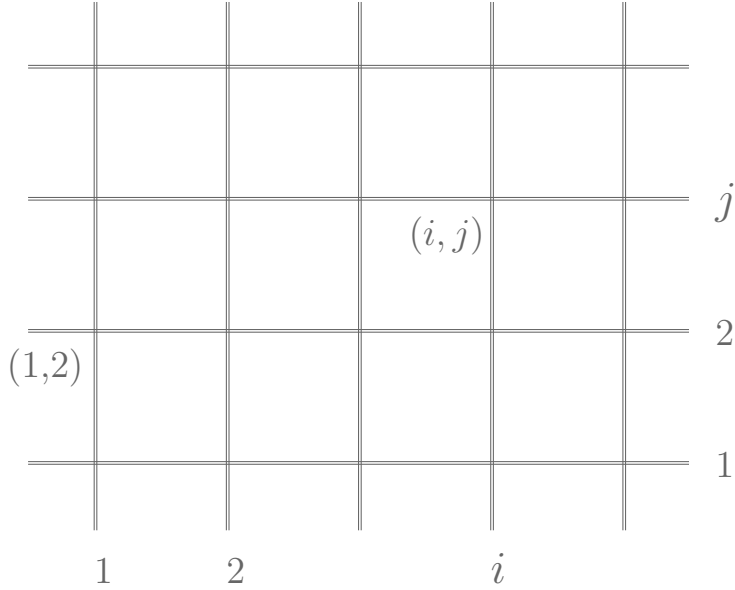

Later that year, Kaufman and 
Onsager{\ccite{KaufmanOnsager1949}}
went on to calculate
some of the two-spin correlations. Let $i$ label the columns of 
the square lattice  (oriented in the usual manner, with axes 
horizontal and vertical),  and $j$ label the rows, as in Figure
\ref{sqlatt}.  Let the spin at site  $(i,j)$ be $\sigma_{i,j}$, with 
values $+1$ and $-1$. Then the 
total energy is 
\bd E \eq -J \sum_{ij} \sigma_{i,j} \sigma_{i,j+1} -J' 
 \sum_{ij} \sigma_{i,j} \sigma_{i+1,j} \ed
 and the partition function is
 \bd Z \eq \sum_{\sigma} \e^{-E/{\kappa T} } \comma \ed
 the sum being over all values of all the spins, 
 $\kappa$ being Boltzmann's constant and $T$ the temperature.
 Onsager defines $H, H', H^*$ by
 \be \label{defH*}
  H = J/{\kappa T} \sep H' = J'/{\kappa T} \sep
  \e^{-2H^*}  = \tanh H \period \ee
  The specific heat diverges logarithmically at a critical 
  temperature $T_c$, where
  \bd \sinh (2J/\kappa T_c ) \sinh (2J'/\kappa T_c )  = 1
   \period \ed
   

 The correlation between the two spins at sites 
 $(1,1)$ and $(i,j)$ is 
 \bd \langle \sigma_{1,1} \sigma_{i,j} \rangle \eq 
 Z^{-1}  \sum_{\sigma}  \sigma_{1,1} 
 \sigma_{i,j} \e^{-E/{\kappa T }} \comma \ed
 For the isotropic case $H' = H$,
Kaufman and Onsager{\ccite{KaufmanOnsager1949}}
give in their equation 43 the formula
 for the correlation
between two spins in the same row:{\bl \footnote{ We have 
negated their $\Sigma_r$, 
which makes them the same as those in Appendix A, and 
corrected what appears to be a sign error.  It is now the same as 
III.43 of the draft paper below.}}
\be \label{III.43}
\langle \sigma_{1,1} \sigma_{1,1+j} \rangle 
\eq   \cosh^2 H^{*} \, \Delta_j  - \sinh^2 H^* \, \Delta_{-j} \ee
Here $\Delta_j$ and $\Delta_{-j}$ are Toeplitz
determinants:
 \bd  \Delta_{j} \eq \raisebox{-0mm}{$
\left| \begin{array}{lllll}
 \Sigma_1 & \Sigma_2 & \Sigma_3  & \cdots & \; \Sigma_{j}  \\
\Sigma_{0} & \Sigma_1 & \Sigma_2  & \cdots & \Sigma_{j-1} \\
    \cdot &     \cdot &     \cdot &     \cdots &     \;  \;   \cdot  \\
\Sigma_{2-j}  & \cdot  & \cdot & \cdots & \; \Sigma_1\end{array}
       \right|  $} \ed
       
\be  \label{Dmk}
\Delta_{-j} \eq \raisebox{-0mm}{$
\left| \begin{array}{lllll}
 \Sigma_{-1} & \Sigma_{-2} & \Sigma_{-3}  & \cdots & \; \Sigma_{-j}  \\
\Sigma_{0} & \Sigma_{-1} & \Sigma_{-2}  & \cdots & \Sigma_{1-j} \\
    \cdot &     \cdot &     \cdot &     \cdots &     \;  \;   \cdot  \\
\Sigma_{j-2}  & \cdot  & \cdot & \cdots & \; \Sigma_{-1}\end{array}
       \right|  $} \ee
where
\bd \Sigma_r   \eq \frac{1}{2 \pi} \int_0^{2 \pi} 
\e^{\iqq  r \omega +\iqq \delta' (\omega)} \, \dv \omega \ed
and 
\be \label{tandelta}
 \tan \delta '(\omega)  \eq \frac{\sinh 2 H \sin \omega }
{\coth 2 H' -\cosh 2 H \cos \omega  } \period \ee
Setting  
\be \label{deldelp}
\delta ( \omega) = \delta'(\omega) + \omega \comma \ee
this implies
\be  \label{eidelta}
 \e^{\iqq  \delta(\omega)  } \eq     \left\{ 
\frac{(1-\coth H' \, \e^{-2 H + \iqq  \omega})(1-\tanh H' \, \e^{-2 H + \iqq  \omega})}
{(1-\coth H' \,  \e^{-2 H - \iqq  \omega})(1-\tanh H' \,  \e^{-2 H - \iqq  \omega})} 
\right\}^{1/2}  \period \ee
(Equations (\ref{tandelta}), (\ref{eidelta}) follow from (89) 
of {\ccite{Onsager1944}}  and are true for the general case when $H'$, $H$
are not necessarily equal.)

Kaufman and Onsager also give the 
formula for the correlation between spins in adjacent rows.
In particular, from their equations 17 and 20, we obtain
 \be \label{1stform}
 \langle \sigma_{1,1} \sigma_{2,2} \rangle \eq 
  \cosh ^2 H^*  \,  \Sigma_1 +   \sinh ^2 H^* \,  \Sigma_{-1}
  \period \ee 

The long-range order, or spontaneous magnetization, can be defined
as
\be \label{defM0}
M_0 \eq \left( \lim_{j \rightarrow \infty}  \langle 
\sigma_{1,1} \sigma_{1,1+j} \rangle  \right) ^{1/2} \period \ee
It is expected to be zero for $T$ above the critical
temperature, and positive below it, as in Figure
 \ref{graphM}.
 
 Kaufman and Onsager were obviously very close to
 calculating $M_0$: all they needed to do was to
 evaluate $\Delta_j$, $\Delta_{-j}$ in the limit $j \rightarrow 
 \infty$. Here I shall present the evidence that they devoted 
their attention to doing so, and indeed succeeded. They used two
methods: the first is discussed in section 2 and Appendix A,
the second in sections 3, 4, 5. Many years ago (probably  
in the early or mid-1990's), John Stephenson (then in Edmonton, 
Canada) sent the author
a photocopy of an eight-page  typescript, bearing the 
names Onsager and  Kaufman,  that deals with the topic. 
Stephenson had copied it about 1965 in Adelaide, Australia,
from  a copy owned by  Ren Potts: both Potts and
Stephenson were students of 
Cyril Domb.{\ccite{Domb1974, Domb1990} }  It's possible that 
Potts's copy had come from Domb when they were together in Oxford 
from 1949 to 1951, but more likely that he had been given it by 
Elliott Montroll, with whom Potts had  collaborated in the early 1960's 
(see ref. {\ccite{MPW1963}}). A transcript of the author's 
copy  is given in  section 3, and a scanned copy 
forms Appendix B.  

Both these methods start from the formulae for the pair correlation
of two spins deep within an infinite lattice. There is also a third method
used by the author for the superintegrable chiral Potts model
(which is an  $N$--state  generalization of the Ising 
model){\ccite{baxter2008}}:
 if one calculates the single-spin expectation 
value $\langle \sigma_{1,M} \rangle$ in a lattice of width $L$ and 
height $2 M$ with cylindrical 
boundary conditions and fixed-spin boundary conditions
on the top and bottom rows, then one can write the result as a determinant
of dimension proportional to $L$. (This method is similar to that of
 Yang.{\ccite{Yang1952}}) The determinant is {\it not}
 Toeplitz, but in the limit $M \rightarrow  \infty$ it is a product of Cauchy
determinants, so can be evaluated directly for finite 
$L$.{\ccite{baxter2010a,baxter2010b}}

 
 \section{The first method}
 \setcounter{equation}{0}
 
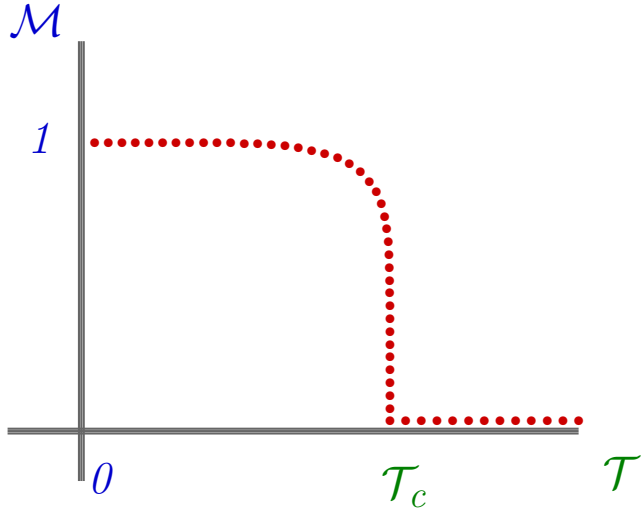
\begin{figure}
\begin{picture}(300,300) (5,-57)
\setlength{\unitlength}{0.9pt}
\thicklines
{\color{black}

\put (110,-35) {\line(0,1) {185}}
\put (111,-35) {\line(0,1) {185}}
\put (112,-35) {\line(0,1) {185}}

\put (80,-15) {\line(1,0) {240}}
\put (80,-14) {\line(1,0) {240}}
\put (80,-13) {\line(1,0) {240}}

{\color{red}

\put(110.0, 105.0)   { \scriptsize $ \bullet $}
\put(115.7, 105.0)   { \scriptsize $ \bullet $}
\put(121.4, 105.0)   { \scriptsize $ \bullet $}
\put(127.1, 105.0)   { \scriptsize $ \bullet $}
\put(132.8, 105.0)   { \scriptsize $ \bullet $}
\put(138.5, 105.0)   { \scriptsize $ \bullet $}
\put(144.2, 105.0)   { \scriptsize $ \bullet $}
\put(149.9, 105.0)   { \scriptsize $ \bullet $}
\put(155.7, 105.0)   { \scriptsize $ \bullet $}
\put(161.4, 105.0)   { \scriptsize $ \bullet $}
\put(167.1, 104.9)   { \scriptsize $ \bullet $}
\put(172.8, 104.7)   { \scriptsize $ \bullet $}
\put(178.5, 104.5)   { \scriptsize $ \bullet $}
\put(184.2, 104.2)   { \scriptsize $ \bullet $}
\put(189.9, 103.7)   { \scriptsize $ \bullet $}
\put(195.5, 102.9)   { \scriptsize $ \bullet $}
\put(201.1, 101.9)   { \scriptsize $ \bullet $}
\put(206.6, 100.5)   { \scriptsize $ \bullet $}
\put(212.0, 98.61)   { \scriptsize $ \bullet $}
\put(217.0, 96.10)   { \scriptsize $ \bullet $}
\put(221.6, 92.87)   { \scriptsize $ \bullet $}
\put(225.4, 88.88)   { \scriptsize $ \bullet $}
\put(228.3, 84.27)   { \scriptsize $ \bullet $}
\put(230.4, 79.27)   { \scriptsize $ \bullet $}
\put(231.8, 74.05)   { \scriptsize $ \bullet $}
\put(232.7, 68.73)   { \scriptsize $ \bullet $}
\put(233.3, 63.37)   { \scriptsize $ \bullet $}
\put(233.6, 58.00)   { \scriptsize $ \bullet $}
\put(233.8, 52.62)   { \scriptsize $ \bullet $}
\put(233.9, 47.23)   { \scriptsize $ \bullet $}
\put(233.9, 41.85)   { \scriptsize $ \bullet $}
\put(234.0, 36.46)   { \scriptsize $ \bullet $}
\put(234.0, 31.08)   { \scriptsize $ \bullet $}
\put(234.0, 25.69)   { \scriptsize $ \bullet $}
\put(234.0, 20.31)   { \scriptsize $ \bullet $}
\put(234.0, 14.92)   { \scriptsize $ \bullet $}
\put(234.0, 9.539)   { \scriptsize $ \bullet $}
\put(234.0, 4.155)   { \scriptsize $ \bullet $}
\put(234.0, -1.230)   { \scriptsize $ \bullet $}
\put(234.0, -6.615)   { \scriptsize $ \bullet $}
\put(234.0, -12.00)   { \scriptsize $ \bullet $}
\put(240.615, -12)   { \scriptsize $ \bullet $}
\put(247.23, -12)   { \scriptsize $ \bullet $}
\put(253.845, -12)   { \scriptsize $ \bullet $}
\put(260.46, -12)   { \scriptsize $ \bullet $}
\put(267.075, -12)   { \scriptsize $ \bullet $}
\put(273.69, -12)   { \scriptsize $ \bullet $}
\put(280.305, -12)   { \scriptsize $ \bullet $}
\put(286.92, -12)   { \scriptsize $ \bullet $}
\put(293.535, -12)   { \scriptsize $ \bullet $}
\put(300.15, -12)   { \scriptsize $ \bullet $}
\put(306.765, -12)   { \scriptsize $ \bullet $}
\put(313.38, -12)   { \scriptsize $ \bullet $}

}

}
{\color{blue}
\put (72,152){\Large  {$\cal   M$}}
  }

{\color{green}
\put (224,-42){\Large  {$\cal  T$}}
\put (231,-46){ \large {\it  c}}
  }

{\color{blue}
\put (71,103){\Large  {\it 1}}
  }

{\color{blue}
\put (92,-40){\Large  {\it 0}}
  }

{\color{green}
\put (304,-38){\Large  {$\cal   T$}}
  }

 \end{picture}
 \caption{ ${\cal M}$ as a function of temperature $\cal T$.}
  \label{graphM}
 \end{figure}

In August 1948, Onsager silenced a conference at Cornell
by writing on the blackboard an exact formula for 
$M_0.${\ccite[p.457] {LongHigginsFisher}}
The following year, in May 1949 at a conference 
of the International Union of Physics on statistical mechanics
in Florence, Italy,{\ccite[p. 261]{Onsager1949}}
Onsager referred to 
the magnetization of the Ising model and announced that
``B. Kaufman and I have recently solved'' this problem.
 He gave  the result as 
\be  \label{Isingmag}
M_0 \eq (1-k^2)^{1/8} \ee
where
\be k = 1/(\sinh 2 H   \,  \sinh 2 H' )  \ee
and the result is true for $0 <k <1$, when $T < T_c$.
 For $k >1$ the magnetization 
vanishes, i.e. ${\cal M} = 0 $. 
Figure \ref{graphM} shows the resulting graph of 
$M_0$ for the isotropic case $H' = H$.

Onsager and Kaufman did not publish their derivation, 
which has led to speculation as to why they did not do so. 
The first published derivation was not until 1952, by 
C. N. Yang,{\ccite{Yang1952}}  who later described the calculation as 
``the longest in my career. Full of  local, tactical tricks, 
the calculation proceeded by twists and 
turns.''{\ccite[p.11]{Yang1983}}

Onsager did outline what happened in an article published in 
1971.{\ccite{Onsager1971a}} He starts by remarking that 
correlations along a diagonal are particularly simple, and 
gives the formula
\be  \label{2ndform}
\langle \sigma_{1,1} \sigma_{m,m} \rangle 
\eq   D_{m-1}  \ee
where $D_m$, like $\Delta_m$, is an 
$m$ by $m$ determinant:
\be  \label{Toeplitzdet}
 D_{m} \eq   \raisebox{-0mm}{$
\left| \begin{array}{lllll}
 c_0 & c_1 & c_2  & \cdots & \; c_{m-1}  \\
c_{-1} & c_0 & c_1  & \cdots & c_{m-2} \\
    \cdot &     \cdot &     \cdot &     \cdots &     \;  \;   \cdot  \\
c_{1-m}  & \cdot  & \cdot & \cdots & \; c_0\end{array}
       \right|  $} \ee
the $c_r$ being the coefficients in the Fourier expansion
\be  \label{genfn}
f(  \omega ) \eq \e^{\iqq {\widehat \delta} (\omega)}  \eq 
\sum_{r=-\infty}^{\infty} c_r \e^{\iqq r \omega} \ee
of the function{\footnote{I write the $\delta$ of {\ccite{Onsager1971a}}
as  $\widehat{\delta}$.}}
\be
\label{newdet}
 \e^{\iqq {\widehat \delta} (\omega)} \eq \left( \frac{1- k \, \e^{\iqq \omega }}
{1- k  \, \e^{ -\iqq \omega } } \right)^{1/2} \period  \ee

It is clear that Onsager knew this when his paper with Kaufman
was submitted in May 1949, because footnote 7 of
{\ccite{KaufmanOnsager1949}}
states that ``It can be shown that $\delta' = \pi/2 -  \omega/2 $ at 
the critical temperature for correlations along a $45^{\circ}$
diagonal of the lattice.'' Indeed, this result does follow immediately 
from  (\ref{newdet}) when $k = 1$, provided we replace 
$\delta$ in (1.4) by $\widehat{\delta}$. Certainly 
(\ref{2ndform}) is true, being
the special case $J_3= v_3 = 0$ of equations (2.4), (5.13), 
(6.10), (6.12)  of Stephenson's pfaffian 
calculation{\ccite{Stephenson1964}} 
of the diagonal correlations of the triangular lattice Ising 
model. The formula (\ref{2ndform})
agrees with  (\ref{1stform})  when $m=2$.

  The fact that the formula
depends on $H, H'$  only via $k$ is a consequence
of the property that the diagonal transfer matrices of two models, 
with different values of $H$ and $H'$, but the same value 
of $k$, commute.{\ccite[\S7.5]{book}}

In {\ccite{Onsager1971a}}, Onsager says that he first evaluated
$D_m$ in the limit $m \rightarrow \infty$ by using generating
functions to calculate the characteristic numbers
(eigenvalues) of the matrix
$D_m$ and that this leads to an integral equation
with a kernel of the form
\be \label{kernel}
 K(u,v) \eq K_1(u+v) + K_2(u-v) 
\period \ee
He then obtained the determinant by taking the product
of the eigenvalues and says that ``This was the basis for 
the first announcements of the result.''. 
 We show how this can be 
done in Appendix A. One does indeed find a kernel of the form
(\ref{kernel}), and go on to obtain
\be \label{detD}
\lim_{m \rightarrow \infty} D_m \eq (1-k^2)^{1/4}
\comma \ee
which is  the result (\ref{Isingmag}) that Onsager announced  
in Cornell in 1948 and  Florence in 1949.

 
 \section{The second method}
 \setcounter{equation}{0}
 
 
 In his 1971 article{\ccite{Onsager1971a}}  Onsager goes on to say that 
 after evaluating the particular determinant $D_{\infty}$ by the
  integral equation method,
 he looked for a method for the evaluation of a
 general infinite-dimensional Toeplitz
 determinant  (\ref{Toeplitzdet}), with arbitrary entries $c_r$.
 (The $c_r$ must tend to zero as $r \rightarrow \pm \infty$ 
 sufficiently fast for the sum in (\ref{genfn}) to be uniformly 
 convergent when $\om $ is real.)
  As soon as he 
 tried rational functions of the form
  \be  \label{genfn3}
f(\omega)  \eq  \frac{\prod (1-\alpha_j \e^{\iqq \omega })}
{\prod (1-\beta_k \e^{-\iqq \omega} )}
\sep  \; \; |\alpha_j|, |\beta_k | <1 \comma \ee
``the general result stared me in the face. Only, before I knew what sort
of conditions to impose on the generating function, we talked to Kakutani
and Kakutani talked to Szeg{\H o}, and the mathematicians
 got there first.''
 
 In another article in the same book {\ccite{Onsager1971b}},
 Onsager gives  further explanation of that comment,
saying that he had found 
``a general formula for the evaluation of Toeplitz 
matrices.{\bl \footnote{Refs. {\ccite{Onsager1971a}},
{\ccite{Onsager1971b}} are reprinted in 
Onsager's collected works, pages {\bl{ 232 -- 241}}
  and {\bl {37 -- 45}}, 
respectively.{\ccite{Onsager1996}} }}
 The only thing I did not know was how to fill out the holes
 in the mathematics and show the epsilons and the deltas 
 and all of that''.  Onsager adds  that six years later
 the mathematician Hirschman told him that he could
 readily have  completed his proof by using
 a theorem of  Wiener's.
 
 There is contemporary evidence to support these statements
 in the form of correspondence in 1950 between Onsager and 
 Kaufman. There is also the photocopy of a 
 typescript mentioned in the Introduction, which deals with 
 the problem of calculating the $\Delta_k$ of 
(\ref{Dmk}).  Here  I present a transcript of it,  with approximately
 the original layout and 
 pagination. A scanned copy is in Appendix B.
 It seems highly likely that this is Kaufman's initial draft of 
 paper IV in their series of papers on Crystal Statistics (see points
 3 and 4 in section 5).
 
 Hand-written additions are shown in red (for contemporary additions)
 and in  blue (for probably later additions). Not all additions 
 are shown.   The  ``Fig. 1'' mentioned on page 2, after equation III.45,
 may be Fig. 4 of  {\ccite{KaufmanOnsager1949}}. Equation (17) 
 on page 7 follows from eqn. 89b of {\ccite{Onsager1944}} after 
 interchanging  $H'$ with $H^*$, $\delta' $ with $\delta^*$ and 
 setting $H' = H$. There is an error in the equation between 
 (20) and (21): the second $e^{-2H}$ in the numerator should
 be  $e^{2H}$, as should the first  $e^{-2H}$ in the  denominator.
 
 \vspace{12cm}
 




\newpage
\begin{figure}[hbt]
\begin{picture}(280,419) (74,250)
\setlength{\unitlength}{1.0pt}
 \includegraphics[height=29.75cm]{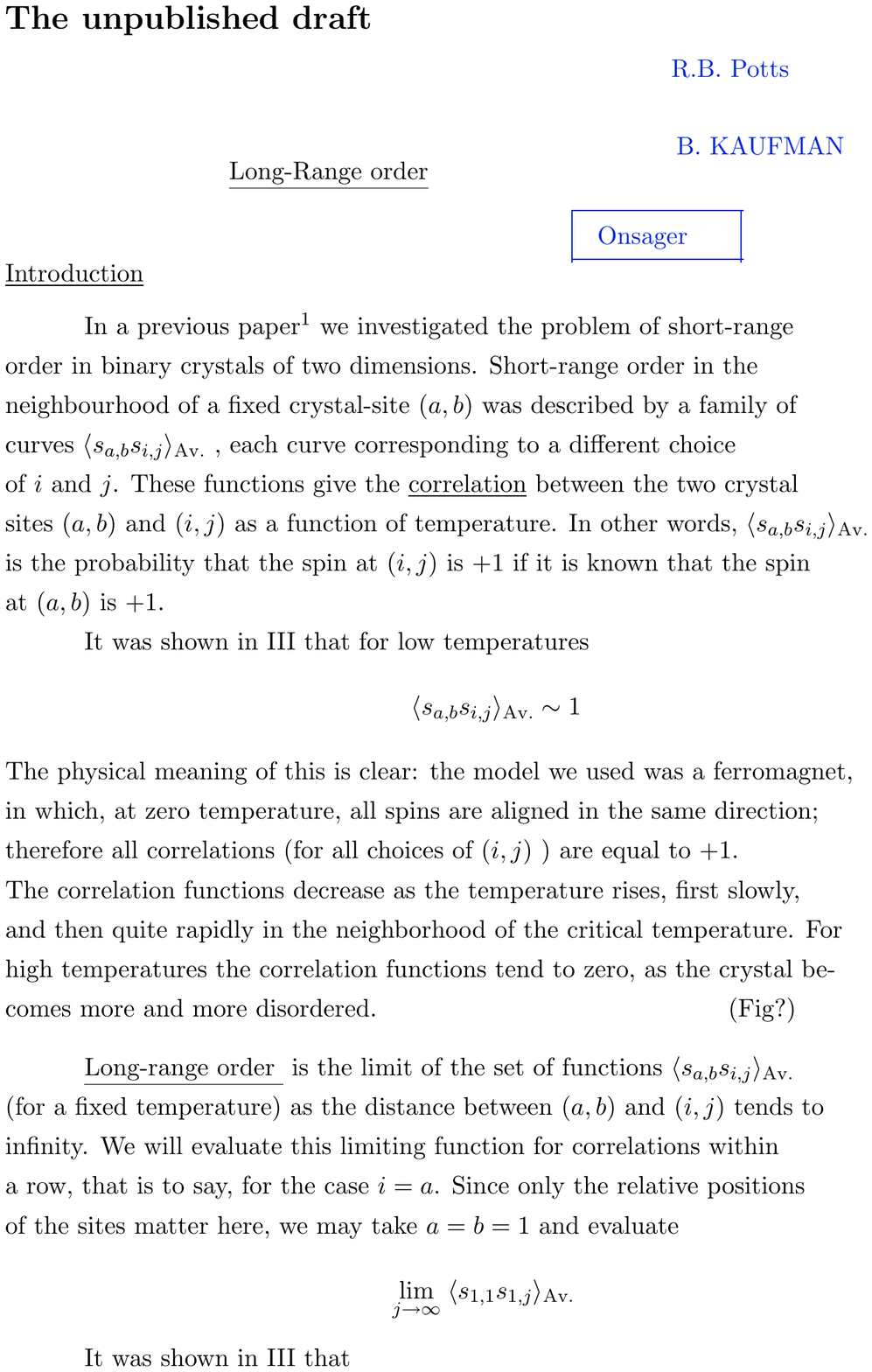}
\end{picture}
 \end{figure}


 \newpage
\begin{figure}[hbt]
\begin{picture}(280,473) (76,280)
\setlength{\unitlength}{1.0pt}
 \includegraphics[height=29.8cm]{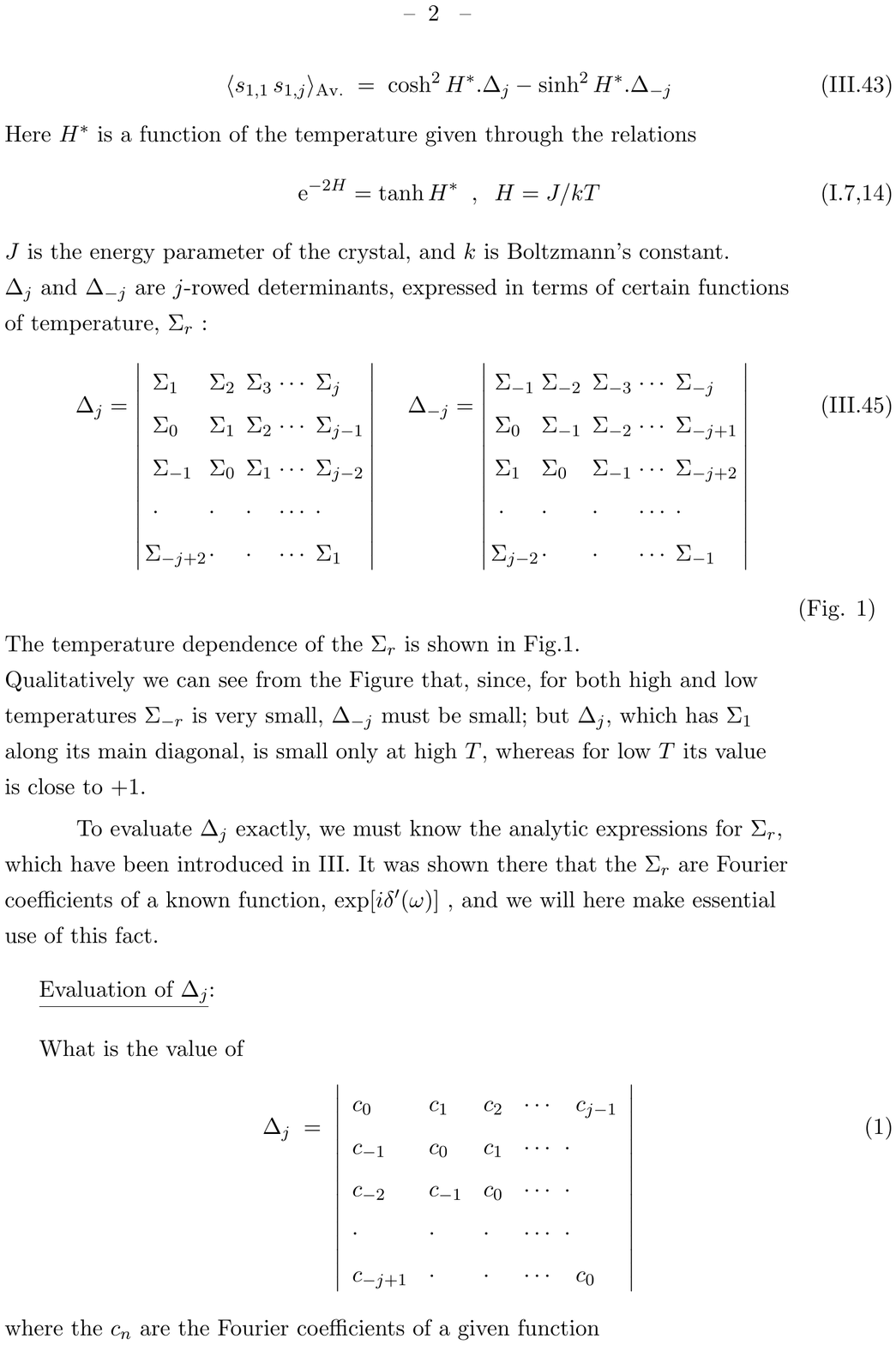}
\end{picture}
 \end{figure}
\newpage


\begin{figure}[hbt]
\begin{picture}(280,426) (70,280)
\setlength{\unitlength}{1.0pt}
 \includegraphics[height=29.2cm]{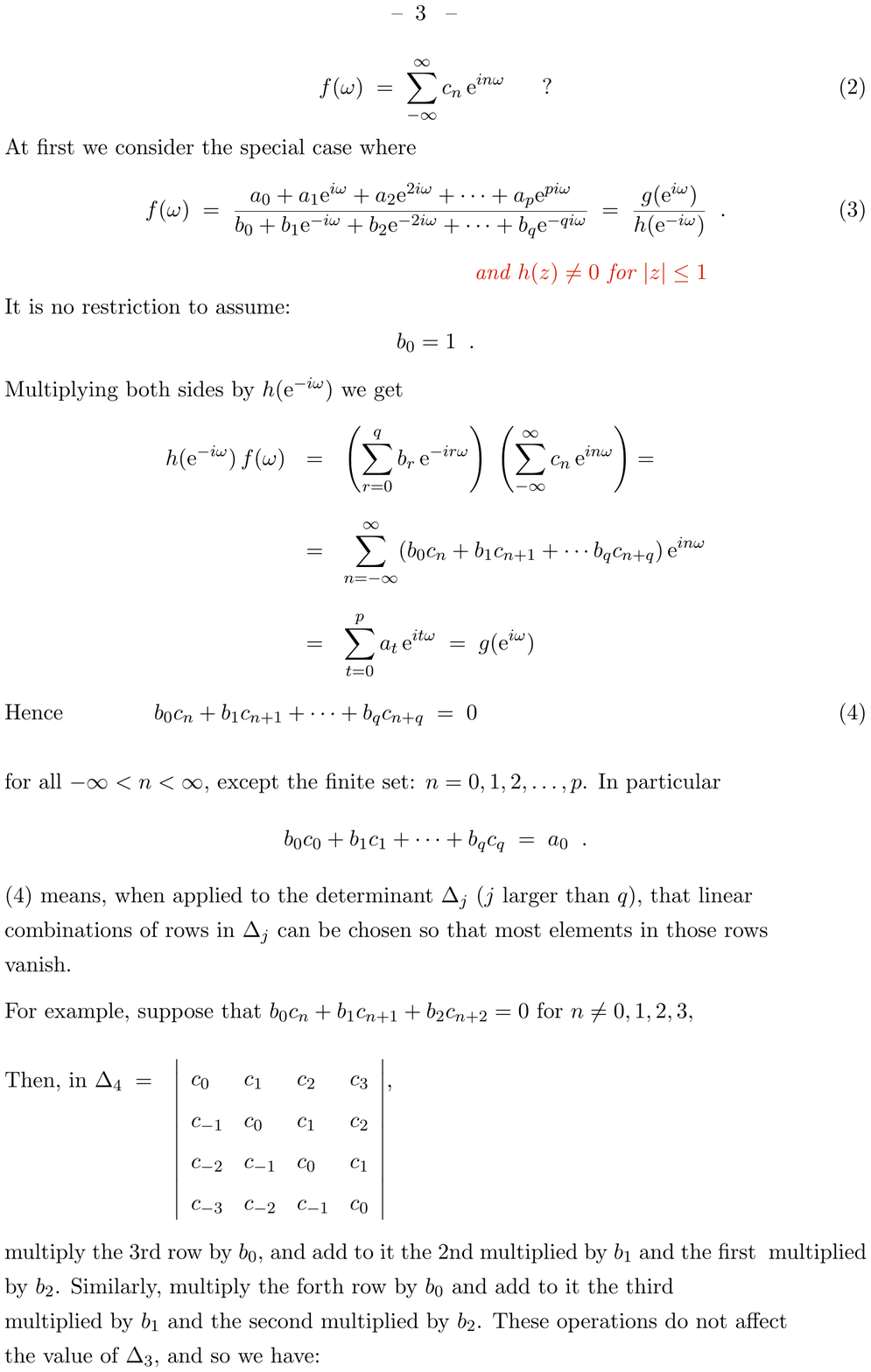}
\end{picture}
 \end{figure}
\newpage


\begin{figure}[hbt]
\begin{picture}(280,416) (74,280)
\setlength{\unitlength}{1.0pt}
 \includegraphics[height=29.8cm]{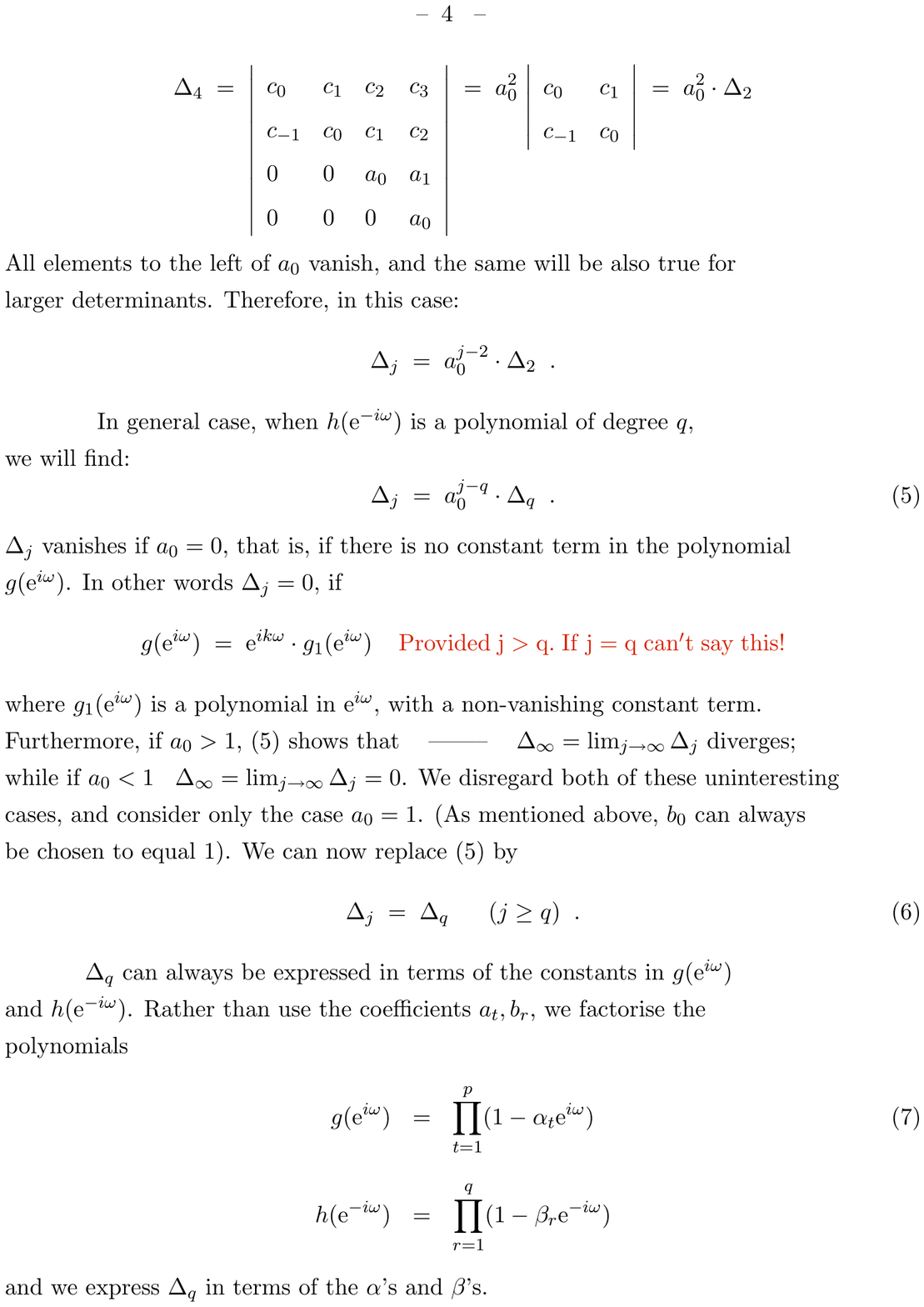}
\end{picture}
 \end{figure}
\newpage
   

\begin{figure}[hbt]
\begin{picture}(280,416) (74,280)
\setlength{\unitlength}{1.0pt}
\includegraphics[height=29.8cm]{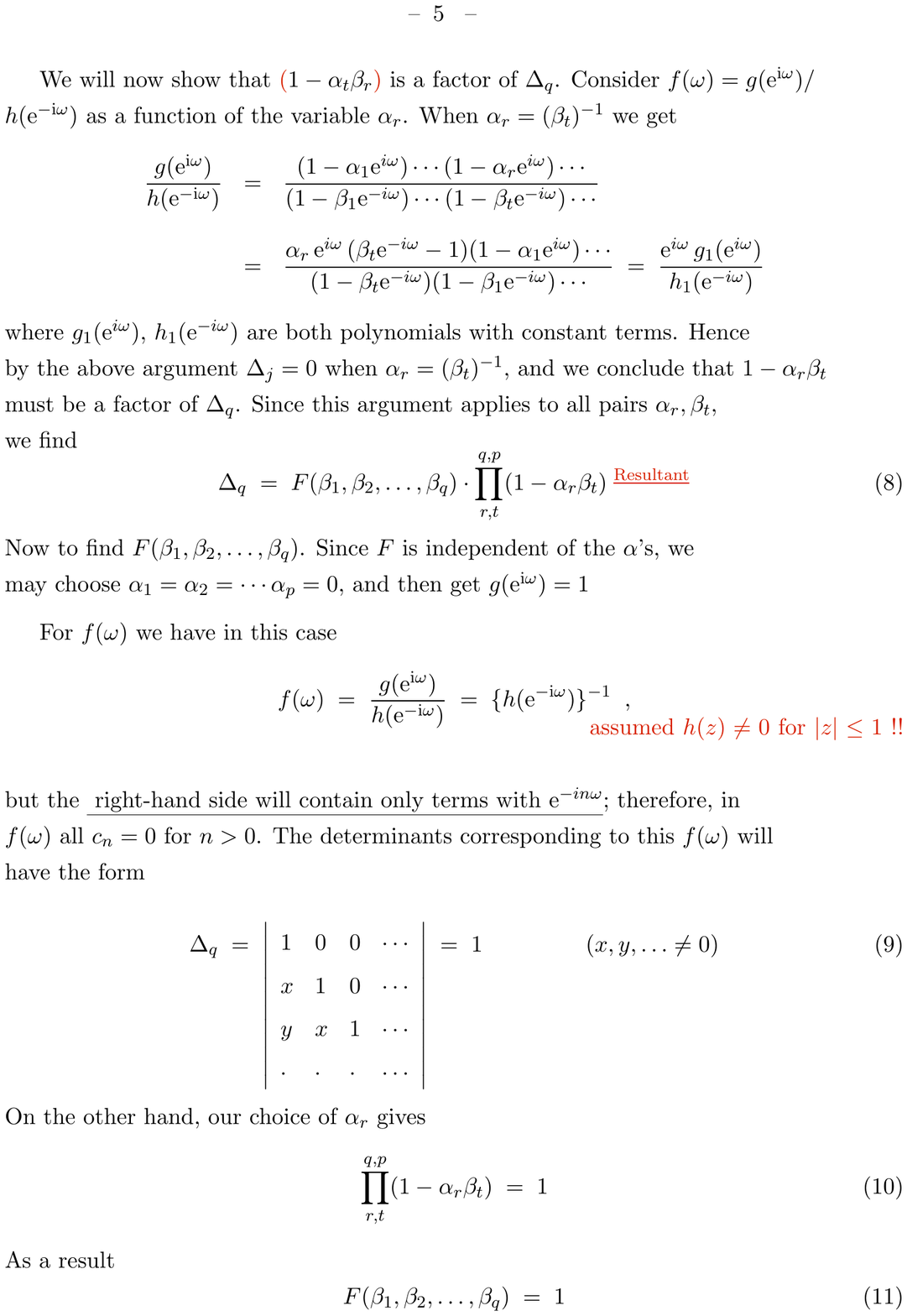}
\end{picture}
 \end{figure}
\newpage


\begin{figure}[hbt]
\begin{picture}(280,416) (74,280)
\setlength{\unitlength}{1.0pt}
\includegraphics[height=29.8cm]{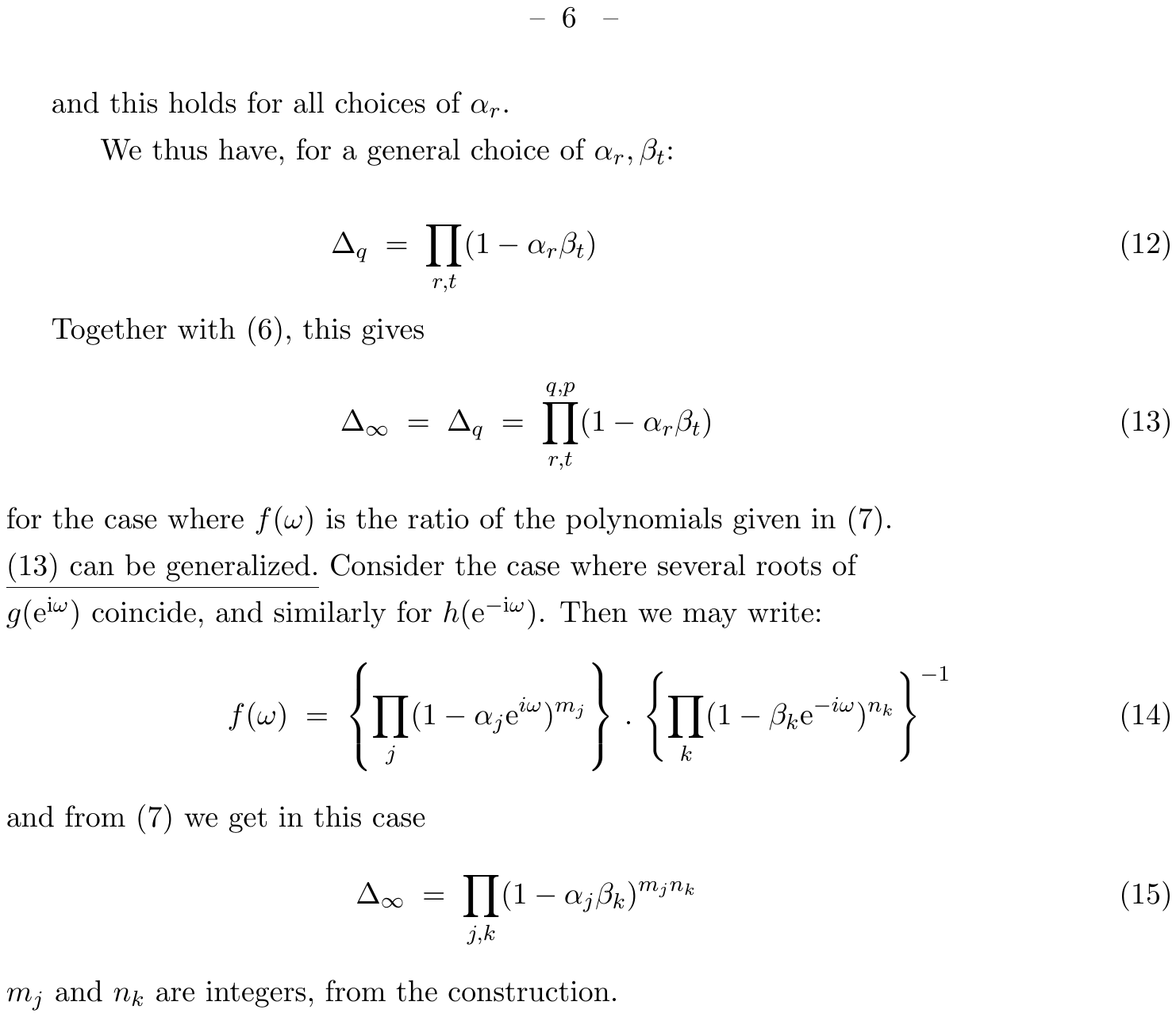}
\end{picture}
\end{figure}
\newpage
 

\begin{figure}[hbt]
\begin{picture}(300,462) (72,245)
\setlength{\unitlength}{1.0pt}
\includegraphics[height=29.2cm]{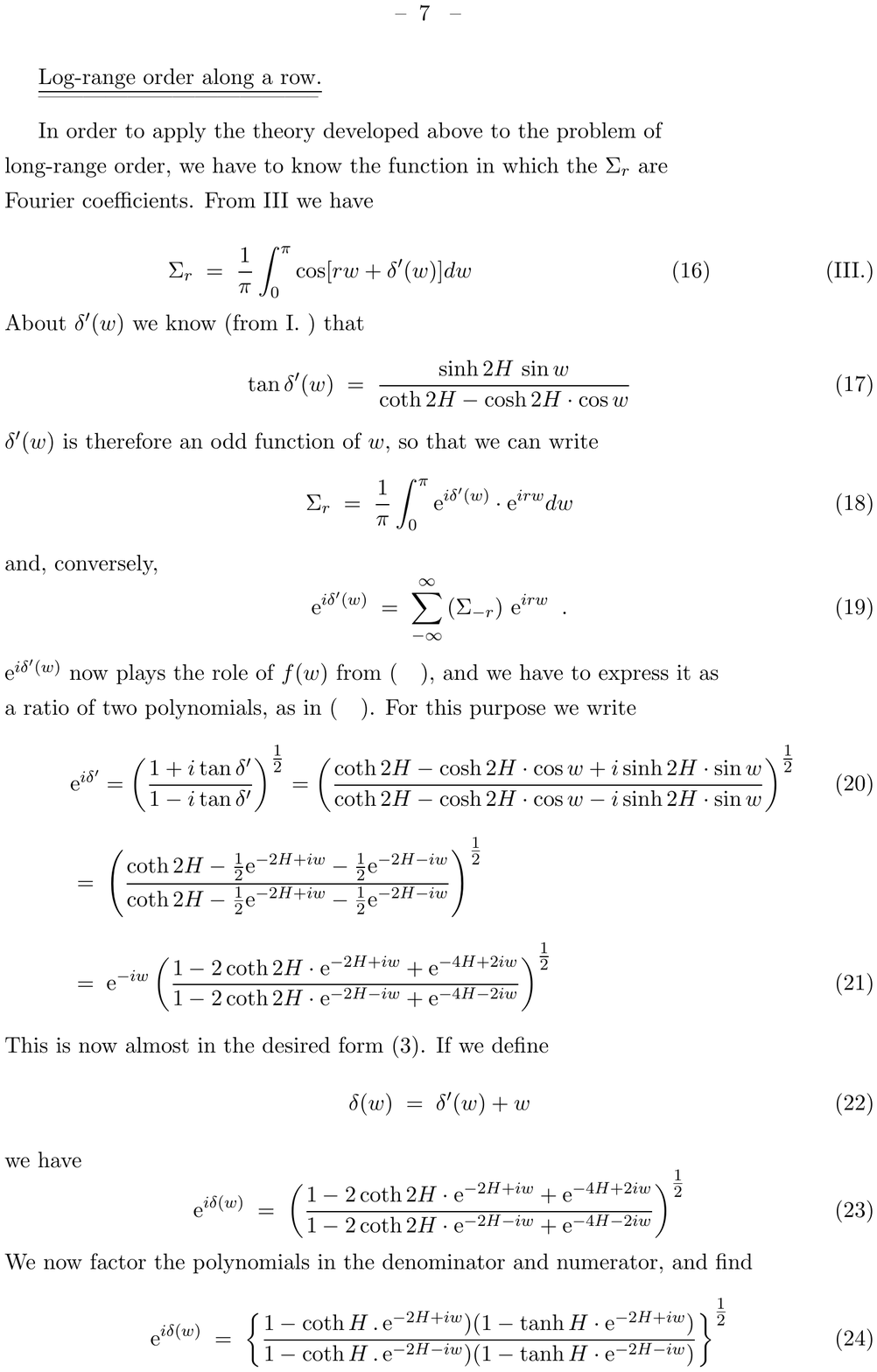}
\end{picture}
\end{figure}
\newpage


\begin{figure}[hbt]
\begin{picture}(260,448) (74,280)
\setlength{\unitlength}{1.0pt}
\includegraphics[height=29.7cm]{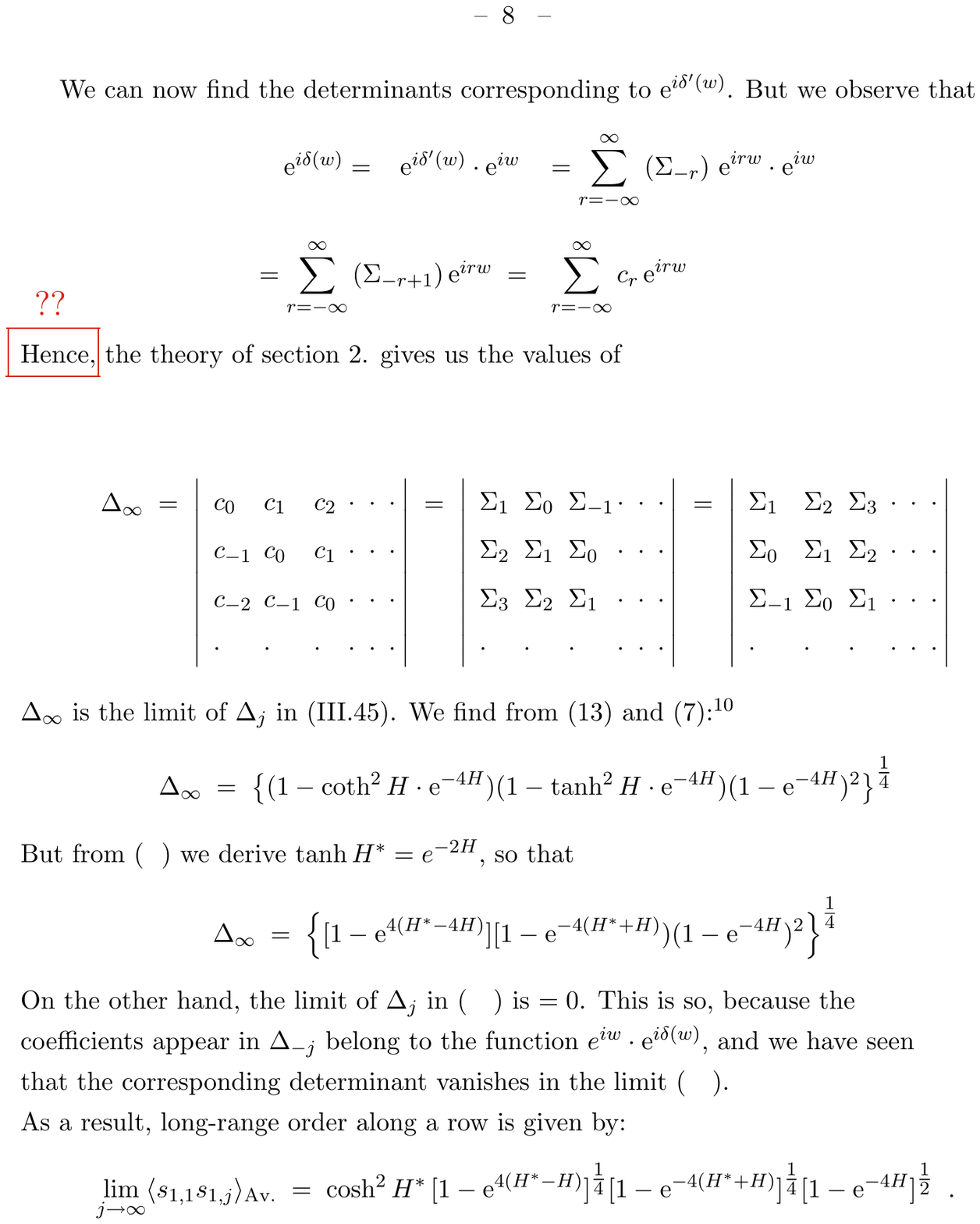}
\end{picture}
\end{figure}
\newpage




\section{Summary of the draft}
\setcounter{equation}{0}
\renewcommand{\theequation}{\arabic{section}.\arabic{equation}}


The names at the top of the first page have been added  by hand 
to the typescript, so it is not immediately obvious who are the 
authors.
However, from the first sentence  and their frequent references to 
paper III , specifically to III.45  (particularly in conjunction with their 
correspondence discussed below) it is clear that the paper is 
by Onsager and Kaufman jointly. Footnote 1 in the first line is not 
given, but is presumably their paper III.

The paper begins by quoting III.43, modified to  (\ref{III.43})
above,  for the isotropic case $H' = H$.

It focusses on the problem of calculating a $j$ by $j$ 
Toeplitz determinant   $\Delta_j$, 
of the general form
(\ref{Toeplitzdet}), in the limit $j \rightarrow \infty$.
The $c_i$ are the coefficients of the 
Fourier expansion (\ref{genfn}) of some function $f( \omega)$, 
initially allowed to be arbitrary.
It first takes $f(\omega)$
\be \label{gform}
 f(\omega ) \eq \frac{a_0 + a_1 \e^{\iqq \omega} + \ldots + 
 a_p \e^{p \iqq \omega}}
{b_0 + b_1 \e^{-\iqq \omega} + \ldots + b_q \e^{-q \iqq \omega}}
\ee
and shows that 
\be \label{vanishes}
\Delta_j =  0 \; \; {\rm if} \; \; a_0 = 0 \; \; {\rm and } \; \; j > q \period \ee

It then takes   $f(\omega)$ to be of the form 
(\ref{genfn3}), or  more specifically
  \be  \label{genfn4}
f(\omega)  \eq  \frac{g(\e^{\iqq \omega})}{h(\e^{-\iqq \omega})}
\comma \ee
where \bd g(\e^{\iqq \omega }) = 
\prod_{t=1}^p (1-\alpha_t \e^{\iqq \omega} )
\sep h(\e^{-\iqq \omega }) = \prod_{r=1}^q (1-\beta_r \e^{-\iqq \omega} )
\comma \ed
and goes on to show, using (\ref{vanishes}), that
\bd \Delta_j \eq \prod_{t=1}^p 
\prod_{r=1}^q  (1- \alpha_t  \beta_r )
 \ed
 provided that  $ j \geq  q$. This is an algebraic identity, true for 
 all
 $ \alpha_t, \beta_r$. It is a trivial generalization then to say that
 if
 \bd
 f(\omega ) = \frac{\prod_{j=1}^p (1-\alpha_j \e^{\iqq \omega} )^{m_j}}
{ \prod_{k=1}^q (1-\beta_k \e^{-\iqq \omega} )^{n_k} }\comma \ed
 then
 \be \label{formoff}  \Delta_\infty  \eq \prod_{j=1}^p
\prod_{k=1}^q  (1- \alpha_j  \beta_k )^{m_j n_k}  \ee
for positive integers $m_j, n_k$.

For $\Delta_j$, $f(\omega) = \e^{\iqq \delta(\omega)}$ is given by
(\ref{eidelta}). 
This is of the general form (\ref{formoff}), with
$p = q=2$ and
\bd \alpha_1 = \beta_1= \coth H \e^{-2H} \sep  \alpha_2 = 
\beta_2 =  \tanh H \e^{-2H}  \sep m_1 = m_2 = n_1 = n_2 = \half 
\ed
so the $m_j, n_j$ are no longer integers. The paper assumes that
 (\ref{formoff}) can be generalized to this case,\footnote{This vital point 
 is considered  by Onsager in the first letter quoted in section 5.} so obtains
 (using $\tanh H = \e^{-2 H^*}$)
  \be \Delta_{\infty} \eq \{ [1 - \e^{4(H^*-H)}][1-\e^{-4(H^*+H)}]
 [1-\e^{-4H}]^2 \}^{1/4} \period  \ee

If we transpose the second determinant (\ref{Dmk}) in 
 (\ref{III.43}), then its generating function is not  $f( \omega)$
 but  $\e^{2\iqq \om }  f(\omega)$. This corresponds to the form
 (\ref{gform}), but with $a_0 = 0$ (and $a_1 = 0$), so $\Delta_{-\infty}$
 should  vanish because of  (\ref{vanishes}).  Thus  (\ref{III.43})
 gives
 \bd \lim_{j \rightarrow \infty} \langle \sigma_{1,1} \sigma_{1, 1+j} \rangle
 \eq \cosh^2 H^* \, \{ [1 - \e^{4(H^*-H)}][1-\e^{-4(H^*+H)}]
 [1-\e^{-4H}]^2 \}^{1/4} \period  \ed
 For the isotropic case $H' = H$, $\e^{-2 H^*} = \tanh H$, so we obtain,
using (\ref{defM0}),
  \bd  M_0^2 \eq \lim_{j \rightarrow \infty} \langle \sigma_{1,1} 
  \sigma_{1, 1+j} \rangle
 \eq  (1 - 1/\sinh^4 2 H)^{1/4}  \eq (1-k^2)^{1/4} \comma \ed
in agreement  with (\ref{Isingmag}).

\subsubsection*{Generalization to the anisotropic case}

In their paper III,{\ccite{KaufmanOnsager1949}}
 Kaufman and Onsager focus on the isotropic case $H' = H$, as 
 does the above draft.  It does in fact appear that  their results 
 generalize immediately to the 
 anisotropic case when $H, H'$ are independent, and 
 $H^*$, $\delta'$ are defined as in {\ccite{Onsager1944}} (i.e. 
 by eqns.  (\ref{defH*}),  (\ref{deldelp}),  (\ref{eidelta}) above),
and  that  (\ref{III.43}), (\ref{1stform}) are then correct as
 written.\footnote{I have verified  that (\ref{1stform}) agrees 
 with (\ref{2ndform}) - (\ref{newdet}). I have not  
 verified   (\ref{III.43})  directly, but
  have checked 
algebraically, with the aid of Mathematica, that  it
agrees with eqn. 56 of {\ccite{MPW1963}} for $ j  \leq 8$.}
 Replacing equation (24) of the draft by  (\ref{eidelta}), and 
 repeating the last few steps, one obtains
  \ba \lim_{j \rightarrow \infty} \langle \sigma_{1,1} 
  \sigma_{1, 1+j} \rangle
 & = & \cosh^2 H^* \, \{ [1 -\coth^2 H' \e^{-4H}][1- \tanh^2 H' \e^{-4H}]
 [1-\e^{-4H}]^2 \}^{1/4} \nonumber \\
  & = &   [1 - 1/(\sinh 2 H \sinh 2 H' )^2 ]^{1/4}   = (1-k^2)^{1/4} 
   \comma \ea
   which again agrees with  (\ref{Isingmag}).


\section{Further comments}
\setcounter{equation}{0}


There are several letters on the Onsager archive in Norway
between Onsager and Kaufman relating to  this second 
method, at

\noindent {\footnotesize 
{\color{\ctnclr}  
\href{http://www.ntnu.no/ub/spesialsamlingene/tekark/tek5/research/009_0097.html}
{http://www.ntnu.no/ub/spesialsamlingene/%
tekark/tek5/research/009{\_}0097.html}}}

and

\noindent {\footnotesize 
{\color{\ctnclr}  
\href{http://www.ntnu.no/ub/spesialsamlingene/tekark/tek5/research/009_0096.html}
{http://www.ntnu.no/ub/spesialsamlingene/%
tekark/tek5/research/009{\_}0096.html}}}

\noindent We shall refer to these two files as 0097 and 0096, respectively.

 \item{{\bf 1.}}  On pages 21 -- 24 of 0097, (also in  {\ccite{JSP1995} })
 is a letter dated April 12, 1950 from 
Onsager to Kaufman giving the argument of the above draft up to
equation 15 therein. Onsager states that the result admits 
considerable 
generalization. He defines
\be    \eta_+  (\omega)   =  \log \gom \sep   \eta_-  (\omega)   = - \log \hom 
\ee
so $f( \om ) =  \e^{  \eta_+  (\omega)  +  \eta_-  (\omega)  }$.
He remarks that if $\log f(\om)$ is analytic in a strip which contains 
the real axis,
then  these functions may be approximated by polynomials which 
have ``no wrong zeros'' in such a manner that the corresponding 
determinants converge. This implies that the $\alpha_j, \beta_j$ all 
have modulus less than one. He  goes on to say that  (15) is equivalent
to
\be  \label{Tform}
\log \Delta_{\infty}  \eq  \frac{\i}{2 \pi} \, \int_{\om = 0}^{2 \pi}  \eta_+ \, \dv  \eta_- ( \om )
 \comma \ee
  If 
  \be \log f( \om ) \eq  \sum_{n=-\infty} ^{\infty} b_n \e^{\iqq n \om } 
 \ee
 and $b_0 = 0$, then (\ref{Tform}) is equivalent to
 \be  \label{Szego}
\log  \Delta_{\infty}  \eq     \sum_{n=1} ^{\infty} n b_n b_{-n}  
\comma \ee
which is the result now known as Szeg{\H o}'s 
theorem.{\ccite{Szego1952,Kac1954,MPW1963}}

Onsager then says ``we get the degree of order from C.S. III without
 much trouble. It equals $(1-k^2)^{1/8}$ as before.''

 \item{\bf 2. }  On pages  {{\bl} 32, 33}  of 0096 is  a letter 
 from Kaufman to Onsager. It  is undated, but was presumably 
 written after April 12th and before April 18th  1950 (the date 
 of the next letter  we discuss). She thanks him for his 
 letter and comments that his method is 
elegant and simple, far superior to the integral-equations 
method for this purpose.

She also mentions the formula (\ref{newdet}), i.e.
\bd f(\om ) \eq \left( \frac{1-k \e^{\iqq \om}}{1-k \e^{-\iqq \om}} \right)^{1/2} 
\ed
for the generating function for long-range order along a diagonal.

 She  goes on to say that the mathematician Kakutani had 
 written to her saying that he had spoken to Onsager about 
 this. He had  been very interested in Onsager's letter, which he 
 saw in her house and immediately copied down.

\item{\bf 3.}  On pages 
{{\bl} 30, 31} of  0096
is a letter from Kaufman to Onsager dated April 18, 1950,
she says she's glad to hear that Onsager is coming to
Princeton the following Friday and refers to Onsager's 1944 
paper I, to her 1949 paper II, to their joint paper III,
and then, significantly to a paper IV ``on long-range order''.
She says she would like Onsager to see her m.s.

\item{\bf 4.}  Then on page  {{\bl} 34} of  0096
is a letter from Kaufman to Onsager dated May 10,
saying ``Here is a draft of Crystal Statistics IV''.

\vspace{4mm}

All this fits with Onsager's recollections of 
1971{\ccite{Onsager1971a, Onsager1971b}}. 
He and Kaufman had evaluated the determinants in 
III.43{\ccite{KaufmanOnsager1949}}, or alternatively in 
(\ref{Toeplitzdet}) - (\ref{newdet}), by the integral equation method
(Appendix A)  and by the general Toeplitz determinant formulae  
(15) of the draft, i.e (\ref{Tform}).
He preferred the second method, but apparently was content to let
Kakutani and Szeg{\H o} take it over and put in the
rigorous mathematics.  Kaufman's draft of paper IV was never 
published, but it seems highly likely that the draft given here
is indeed that.

Szeg{\H o} did publish his resulting
general theorem{\ccite{Szego1952}},
{\ccite[p.76]{Kac1954}}  
on the large-size limit of a Toeplitz 
determinant, but not until 1952. He also restricted his attention 
 to Hermitian forms, where 
 $f^* (\om ) = f(\om )$ and
 $b^*_n = b_{-n}$,{\ccite[footnote 17] {MPW1963}} so his 
 result needed  further generalization before it could be 
 applied to  to the Ising model magnetization.

The first derivation of $M_0$  published  was in 1952 by  C.N. 
Yang.{\ccite{Yang1952} } He used the spinor operator algebra
to write $M_0$
as the determinant of an $L$ by $L$ matrix and evaluated the 
determinant by calculating the eigenvalues of the matrix in 
the limit $L \rightarrow \infty $. Intriguingly, he mentions Onsager and 
Kaufman's papers I, II and III, and in footnote 10 of his paper, 
Yang thanks Kaufman for showing him her notes on Onsager's work.
However, his method is quite different from theirs.


Later, combinatorial ways were found of writing the partition function 
of the Ising model on a finite lattice directly as  a determinant or a 
pfaffian (the square root of an anti-symmetric
determinant).{\ccite{KacWard, HurstGreen}}. Then it was
realised that the problem could be solved by first expressing it 
as one of filling a planar lattice with 
dimers.{\ccite{Kasteleyn1961, Fisher,TemperleyFisher,
Kasteleyn1963}}
In 1963 Montroll, Potts and Ward{\ccite{MPW1963}} used 
these pfaffian 
methods to show that 
$\langle \sigma_{1,1} \sigma_{1,j+1} \rangle $ could be written as a 
single Toeplitz determinant.
They  evaluated its value in the limit 
$j \rightarrow \infty $ limit  by using  Szeg{\H o}'s  theorem.

So there is no reason to doubt that Onsager and Kaufman 
had derived
the formula (\ref{Isingmag}) by May 1949, and ample evidence that 
they had obtained the result by what is now known as Szeg{\H o}'s 
theorem by May 1950. They did not publish the calculation,
perhaps  because
 the mathematicians beat them to the remaining problem of 
``how to fill out the holes
 in the mathematics and show the epsilons and the deltas 
 and all of that''.

 \section{Acknowledgements}
 
 The author is most grateful to John Stephenson for giving
 him a copy of  the draft paper reproduced here, and 
 thanks  Jacques Perk for telling him of 
 ref.{\ccite{JSP1995}} and  the material on 
the Lars Onsager Online archive  under ``Selected research material 
and writings''  at 

\noindent {\footnotesize 
{\color{\ctnclr}  
\href{http://www.ntnu.no/ub/spesialsamlingene/tekark/tek5/arkiv5.php}
{http://www.ntnu.no/ub/spesialsamlingene/tekark/tek5/arkiv5.php}}}
 
\noindent in particular items 9.96 and 9.97, and for pointing out a number of 
typographical errors in the original form of this paper.
He also thanks Harold Widom for sending him a copy of the letter 
 {\ccite{JSP1995}}, and Richard Askey for alerting him to
 page 41 of Onsager's collected works. He is grateful to the reviewers 
 for a number of  helpful comments.




\section*{Appendix A: An integral equation method}
\setcounter{equation}{0}
\renewcommand{\theequation}{A\arabic{equation}}

We regard $D= D_m$ as a matrix and
write $z$ for $\e^{\iqq \omega }$ and $C(z)$ for 
$\e^{\iqq {\widehat \delta} ( \omega)}$, so (\ref{genfn}), (\ref{newdet}) become
\be  \label{defC}
 C(z) \eq  { \left( \frac {1-k z}{1-k/z}  \right) }^{1/2}    =  
 \sum_{m=-\infty}^{\infty} c_r  z^r  \period \ee

For  $0 < k < 1$, the determinant of $D_{m}$ tends to a non-zero limit as 
$m \rightarrow \infty$. 
The eigenvalues of $D$ itself lie on an arc in the
complex plane, and appear to tend to a continuous distribution as
$m \rightarrow \infty $. By contrast, in this limit the individual eigenvalues 
$\lambda_r$ of $D^T D$ occur in discrete pairs, lying on the real axis, 
between 0 and 1. If one orders them so that 
$\lambda_r \leq  \lambda_{r+1}$, then for given $r$ 
the eigenvalue   $\lambda_r$ appears to tend to a limit as 
$m \rightarrow \infty$, and these limiting values tend to 
one as $r \rightarrow \infty$.

We assume these properties and seek to calculate the 
eigenvalues of $D^T D$ in the limit $m \rightarrow \infty $, 
and hence the determinant of  $D^T D$, which is
$(\det D)^2$.
Writing them as $\lambda^2$ , we can write the 
eigenvalue equation as
\be \label{eigveceqn}
\lambda x = D y \sep \lambda y = D^T x \period \ee

For finite $m$,  $x = \{ x_0, x_1, \ldots,  x_{m-1} \}$,
$y = \{ y_0, y_1, \ldots,  y_{m-1} \}$. Let $P$ be the $m$ by 
$m$ matrix with entries
\bd P_{ij} \eq  \delta_{i, m-1-j} \ed 
so $P x =  \{ x_{m-1}, x_{m-2}, \ldots,  x_{0} \}$. Then
\bd   D^T  =  PDP \sep  D^T D = (PD)^2 \period \ed
For finite $m$, the eigenvalues of $D^T D$ are distinct,
so the eigenvectors are those of $PD$.  For large $m$ 
and a given $\lambda$, this means that the elements 
$x_i, y_i$ are of order one if $i$ is close to zero or 
to $m-1$, but tend to zero  in between, when both 
$i$ and $m-i$ become large.

However, in the limit $m \rightarrow \infty$, the eigenvalues
occur in equal pairs. One can then choose the two 
corresponding eigenvectors  so that one has the property
that
\be \label{1sttype}
x_i, y_i  \rightarrow  0 \; \;  \; {\rm as \; \;  } i \rightarrow \infty \comma
\ee
while for  the other eigenvector  $ x_{m-i} , y_{m-i}  \rightarrow  0 $.
The eigenvectors are transformed one to another by 
replacing $x_i$ and $y_i$ by $y_{m-1-i}$ and $x_{m-1-i}$, 
respectively.

Since the eigenvalues are equal, we can and do restrict our
attention to  eigenvectors with the property (\ref{1sttype}).


\subsubsection*{Generating functions}


Taking the limit $m \rightarrow \infty$, we can write 
(\ref{eigveceqn}) explicitly as
\be \label{eigeqn}
 \lambda \, x_i = \sum_{j=0}^{\infty} c_{j-i}  y_j \sep 
\lambda \, y_i = \sum_{j=0}^{\infty} c_{i-j}  x_j   \ee
for $i = 0, 1,2,  \ldots \, $.


We can extend these equations to  negative $i$, defining
$x_i, y_i$ to then be given by the left-hand sides of the
 equations. Let $X(z), Y(z)$ be the generating functions
\bd X(z) = \sum_{i=-\infty}^{\infty} x_i z^i \sep 
 Y(z) = \sum_{i=-\infty}^{\infty} y_i z^i \period \ed
 Then (\ref{eigeqn}) gives
 \be  \label{inteqns}
\lambda \, X(z) = C(1/z) \tilde{Y}(z) \sep
 \lambda \, Y(z) = C(z) \tilde{X}(z) \comma \ee
 where
 \bd \tilde{X}(z) = \sum_{i=0}^{\infty} x_i z^i \sep 
 \tilde{Y}(z) = \sum_{i=0}^{\infty} y_i z^i \period \ed
 
 A Wiener--Hopf argument (or simple contour integration on each
  term)  gives, for $|z| < 1 $,
  \be \label{WH}
  \tilde{X}(z) =  \frac{1}{2 \pi \iqq } \oint \frac{X(w) }{w-z} \, \dv w 
  \sep  \tilde{Y}(z) =  \frac{1}{2 \pi \iqq } \oint \frac{Y(w) }{w-z} \, \dv w  
   \comma \ee
the integrations being round the unit circle in the complex 
$w$-plane.

Thus (\ref{inteqns}) is a pair of coupled integral equations
for $\lambda, X(z), Y(z)$.


\subsubsection*{The integral equations in terms of elliptic
functions}


To get rid of the square root in (\ref{defC}) we introduce
Jacobi elliptic functions of modulus $k$ and set
\bd   z = k \, \sn^2 u \period \ed
If  $K, K'$ are the complete elliptic integrals and 
$ \label{defalpha}
u \eq \alpha - K -\iqq K'/2 $,
then from (8.181), (8.191) of {\ccite{GR}},
 or (15.1.5), (15.1.6) of {\ccite{book}},
\bd   z \eq  s \, \prod_{n=1}^{\infty} 
\frac{(1+p^{4n-3}/s)^2(1+p^{4n-1}s)^2}
{(1+p^{4n-3} \, s)^2(1+p^{4n-1}/s)^2} \ed
where 
\bd p = q^{1/2} = \e^{-\pi K'/2 K } \sep s = \e^{\iqq \pi \alpha/K } 
\sep 0 < q < 1 \period \ed

If $ 0 < k <1$, then $0 < p,q < 1$, so we see that
$z$ goes round the unit circle as $\alpha$ goes from 
$- K $ to $K$ along the real axis, i.e.
$u$ goes along a horizontal line in the complex plane
from   $-2K -\iqq K'/2$ to $-\iqq K'/2$.

From the elliptic function relations
$ \cn^2 u = 1 - \sn^2 u$, $\dn^2 u = 1 - k^2 \sn^2 u$,
\ba  \label{fnC}
C(z)  & = &  \iqq \, \frac{\sn \,  u  \, \dn \, u}{\cn \, u} \comma 
\nonumber \\
& = &  \prod_{n=1} ^{\infty}
 \frac{(1+(-1)^n p^{2n-1} s)(1-(-1)^n p^{2n-1}/s)} 
{(1 -(-1)^n  p^{2n-1} s)(1 +(-1)^n  p^{2n-1}/s)} 
 \ea

Replace $w$ in (\ref{WH}) by
\bd w = k \, \sn^2 v  \comma \ed
where 
\be \label{domain}
{\rm Im} (v) = -K'/2 \sep   -K'/2 < {\rm Im} (u) < 
K'/2 \period \ee
Then
\be \dv w  \eq 2 k \, \sn \, v \, \cn \, v \, \dn \, v  \, \dv v \period \ee

It is helpful to set
\be \label{XYhat}
X(u) \eq \frac{- \iqq \widehat{X} (u)}{\sn \, u\, \dn \, u } \sep 
Y(u) \eq \frac{ \iqq \widehat{Y} (u)}{\cn \, u } \period \ee

Then the integral equations (\ref{inteqns}) become
\be \label{inteq2}
\lambda \widehat{X} (u) =  \frac{1}{2\pi} \int  M(u,v) 
\widehat{Y} (v) \dv v \sep 
\lambda \widehat{Y} (u) = \frac{1}{2\pi}   \int  M(v,u) 
\widehat{X} (v) \dv v 
\comma \ee
where the integrations are from  $-2K -\iqq K'/2$ to $-\iqq K'/2$
and
\bd M(u,v) \eq  2   \, \frac{\cn \, u\, \sn \, v \, \dn \, v }
{\sn^2 \, v  - \sn^2 \, u} \period \ed
Using the Liouville theorem-type arguments of sections 
15.3, 15.4 of {\ccite{book}}, one can establish that
\bd M(u,v) \eq    \frac{\dn  (v+u)}{\sn (v+u) } + 
\frac{\dn  (v-u)}{\sn (v-u) }  \period \ed
We see that this kernel is indeed of the form (\ref{kernel}).

\vspace{2mm}

\setlength{\jot}{4mm}

We require the functions  $\tilde{X}(z), \tilde{Y}(z)$ to be
analytic for $|z| <1$, i.e. for  $|{\rm Im }(u) | <K'/2$, and in
particular for real $u$.  From (\ref{inteqns}), (\ref{fnC}) and 
 (\ref{XYhat}), 
 $ \tilde{X}(z) = \lambda  \widehat{Y}(u)/\sn u \, \dn u$,
 $ \tilde{Y}(z) = \lambda  \widehat{X}(u)/\cn u$.
Since  $\sn \, 0 = \cn \, K = 0$, it follows
 that
 \be \label{restrict}
 \widehat{X}(K) = \widehat{Y}(0) = 0 \period \ee

Also, $M(u,v)$ is an analytic function of 
$u$  within the domain   (\ref{domain}), and we can negate
$u$ within the domain. Since $M(u,v)$ is an 
even function of $u$, it follows from  (\ref{inteq2}) 
(provided $\lambda$ is not zero)
that  $\widehat{X}(u)$ is also an even function.
Similarly, $\widehat{Y}(u)$ is an odd function.


\subsubsection*{Solution by Fourier transforms}


The  sum and difference form of the kernel  ${M}(\alpha, \beta ) $
suggests solving (\ref{inteq2}) by Fourier transforms.
The functions $\widehat{X}, \widehat{Y},  M$ are anti-periodic of 
period $2 K$, while $\widehat{X}(u)$, $\widehat{Y}(u)$
are even and odd, respectively. We therefore try
\be \label{guess}
 {\widehat{X}}(u)    =  A  \cos \pi (2r-1) u/2K  \sep 
   {\widehat{Y}}(u)  =   B \sin  \pi (2r-1) u/2K  \comma  \ee
$r$ being a positive integer. We note that this ansatz immediately 
satisfies  (\ref{restrict}).

Integrating round a period rectangle and using Cauchy's theorem,
 or using (8.146) of {\ccite{GR}}, for all integers $r$ we obtain
\setlength{\jot}{5mm}
\be  \frac{1 }{2 \pi   } \, \int   \frac{\dn (v-u)}
{\sn (v-u)}  \, \e^{\iqq \pi (2r-1) v/2K}  \,  \dv v  =   
 \iqq  \, \frac{\e^{\iqq \pi (2r-1) u/2K}}{1+q^{2r-1}}   \period \ee
 Negating $2r-1$ and/or  $u$ (but {\em not} $v$) and
 taking appropriate sums and differences, we obtain 
 the identities, true when both $v_0+u$ and $v_0-u$ have 
 imaginary parts between $-2K'$ and zero and the integration 
 is from
$v_0 - 2K$ to $v_0$ :
\be \frac{1}{2 \pi} \int M(u,v) \sin {\textstyle \frac {\pi (2r-1) v}{2K} }
 \dv v =  \frac{1-q^{2r-1}}{1+q^{2r-1}} \, 
 \cos {\textstyle \frac {\pi (2r-1) u}{2K} }  \ee
 
 \be \frac{1}{2 \pi} \int M(v,u) \cos {\textstyle \frac {\pi (2r-1) v}{2K} }
 \dv v =  \frac{1-q^{2r-1}}{1+q^{2r-1}} \, 
 \sin {\textstyle \frac {\pi (2r-1) u}{2K} }  \period 
 \ee
 
 We see that the ansatz  (\ref{guess}) does indeed satisfy
 (\ref{inteq2}), with $A = B= 1$ and
 \be \label{lambda}
  \lambda =   \frac{1-q^{2r-1}}{1+q^{2r-1}}  
 \period \ee
 
 More generally, we can allow $\widehat{X}(u)$, 
 $\widehat{Y}(u)$ each to be an arbitrary 
 linear combination  of $\exp[\iqq \pi (2r-1) u/2K]$ and
 $\exp[{ - \iqq \pi (2r-1) u/2K}]$. We obtain the above solution, plus
 another where $\lambda = 0$. However, this 
 second solution does not satisfy
 the necessary condition  (\ref{restrict}).
 


 \subsubsection*{Determinant of  $D$}
 

 For every positive integer $r$ there is one and only one
 eigenvector  of $D^T D$ of the form (\ref{1sttype}), with 
 eigenvalue $\lambda^2$, $\lambda$ being given by
 (\ref{lambda}). However, from the discussion
 before  (\ref{1sttype}), there is also an equal
 eigenvalue  whose eigenvector has elements 
 $x_i, y_i$ of order 1 only when $i$ is close to $m$. Hence,
 using  (8.197.4) of {\ccite{GR}} or
  (15.1.4b) of  {\ccite{book}},
 \be \det D^T  D =   ( \det D )^2 = \prod_{r=1}^{\infty} 
  \left(   \frac{1-q^{2r-1}}{1+q^{2r-1}}   \right)^4  = k'
  = (1-k^2)^{1/2}  \period \ee
   so 
  \be \det D  \eq  (1-k^2)^{1/4}  \ee
  as in (\ref{detD}).
  
 \newpage

\section*{Appendix B: The typewritten draft}
\setcounter{equation}{0}
\renewcommand{\theequation}{B\arabic{equation}}

 \begin{figure}[hbt]
\begin{picture}(320,340) (14,355)
\setlength{\unitlength}{1.0pt}
 \includegraphics[height=25cm]{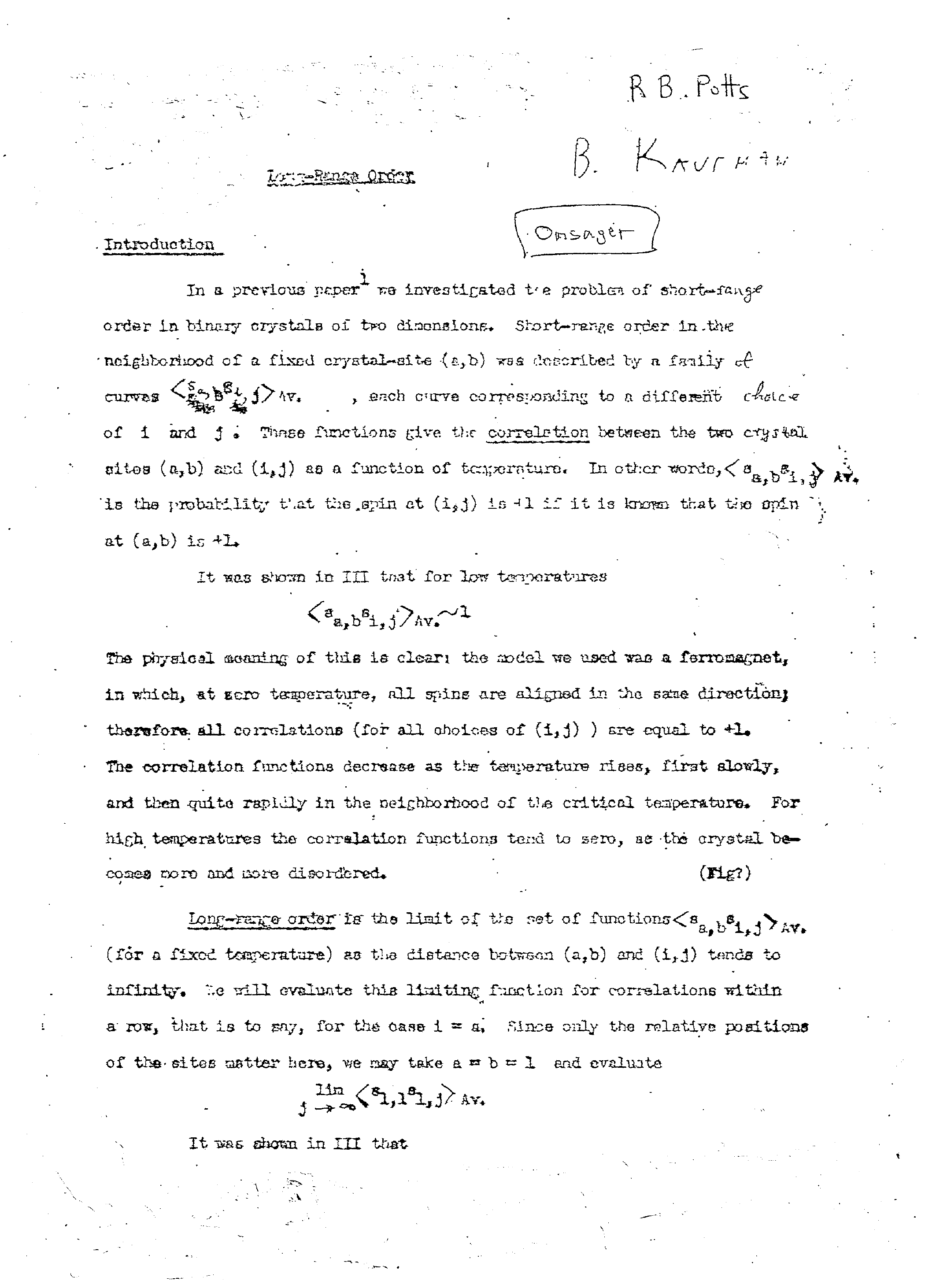}
\end{picture}
 \end{figure}
 \newpage

 \begin{figure}[hbt]
\begin{picture}(320,470) (14,200)
\setlength{\unitlength}{1.0pt}
 \includegraphics[height=25cm]{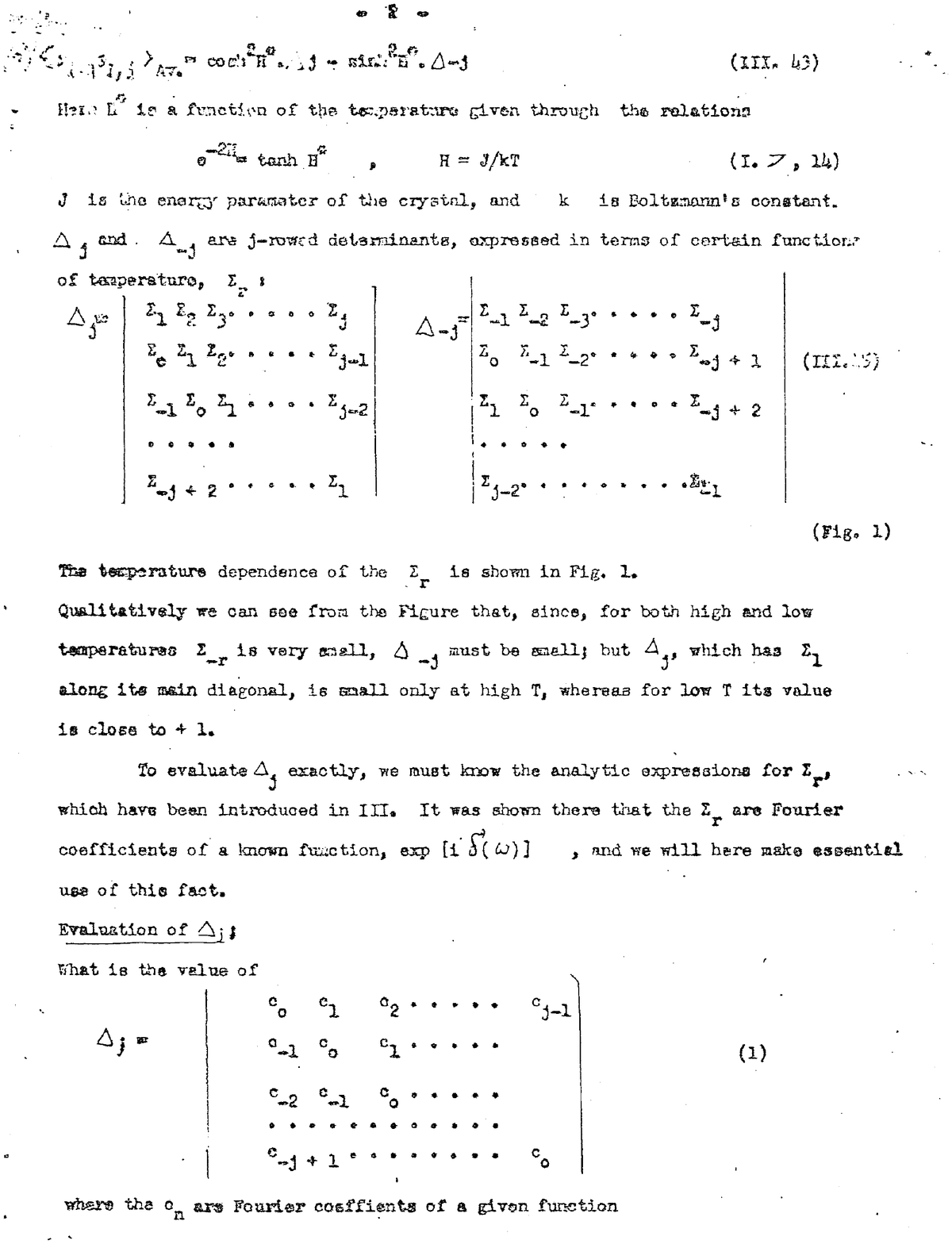}
\end{picture}
 \end{figure}
\newpage

\begin{figure}[hbt]
\begin{picture}(320,470) (14,200)
\setlength{\unitlength}{1.0pt}
 \includegraphics[height=25cm]{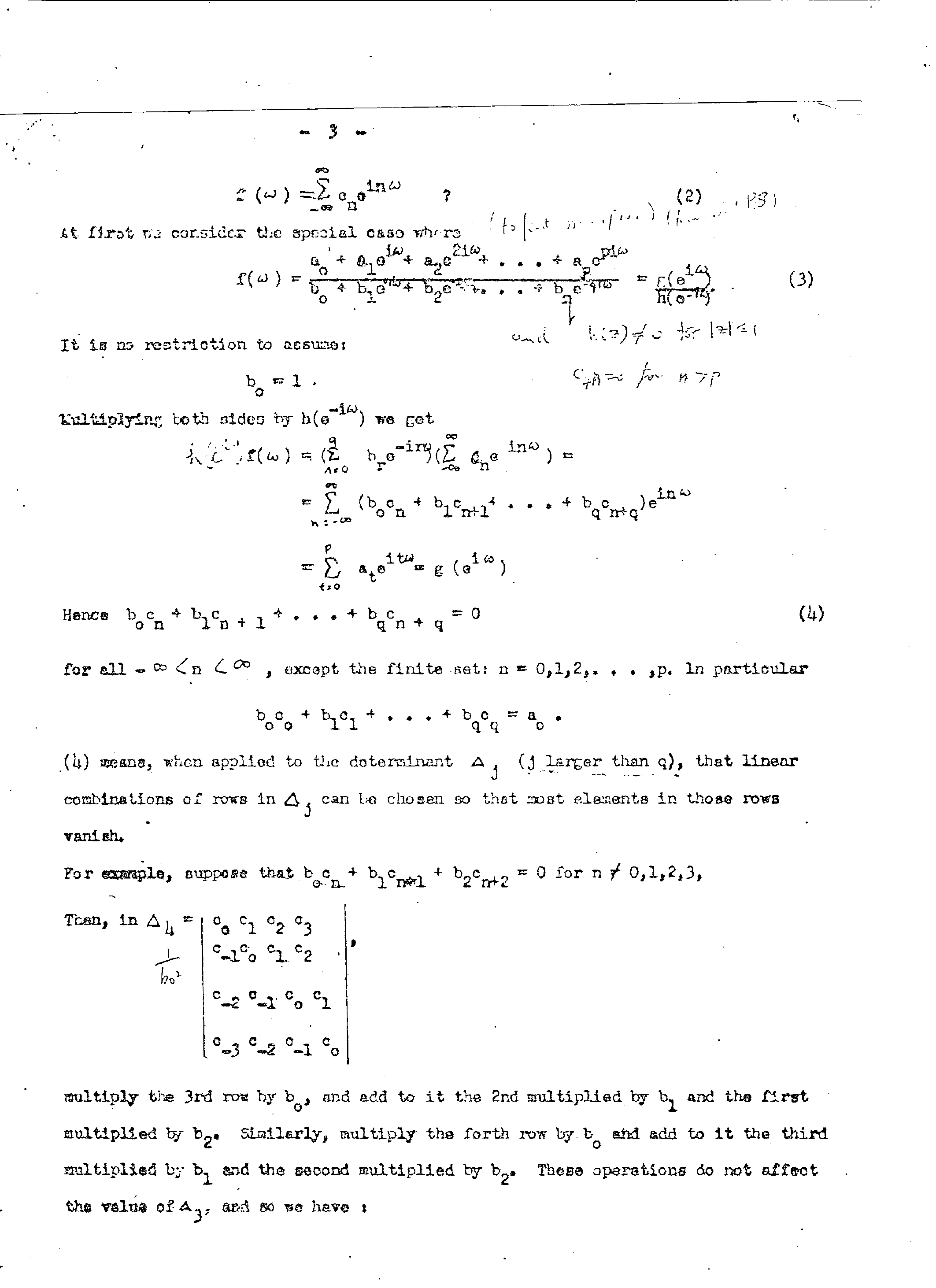}
\end{picture}
 \end{figure}
\newpage

\begin{figure}[hbt]
\begin{picture}(320,470) (14,200)
\setlength{\unitlength}{1.0pt}
 \includegraphics[height=25cm]{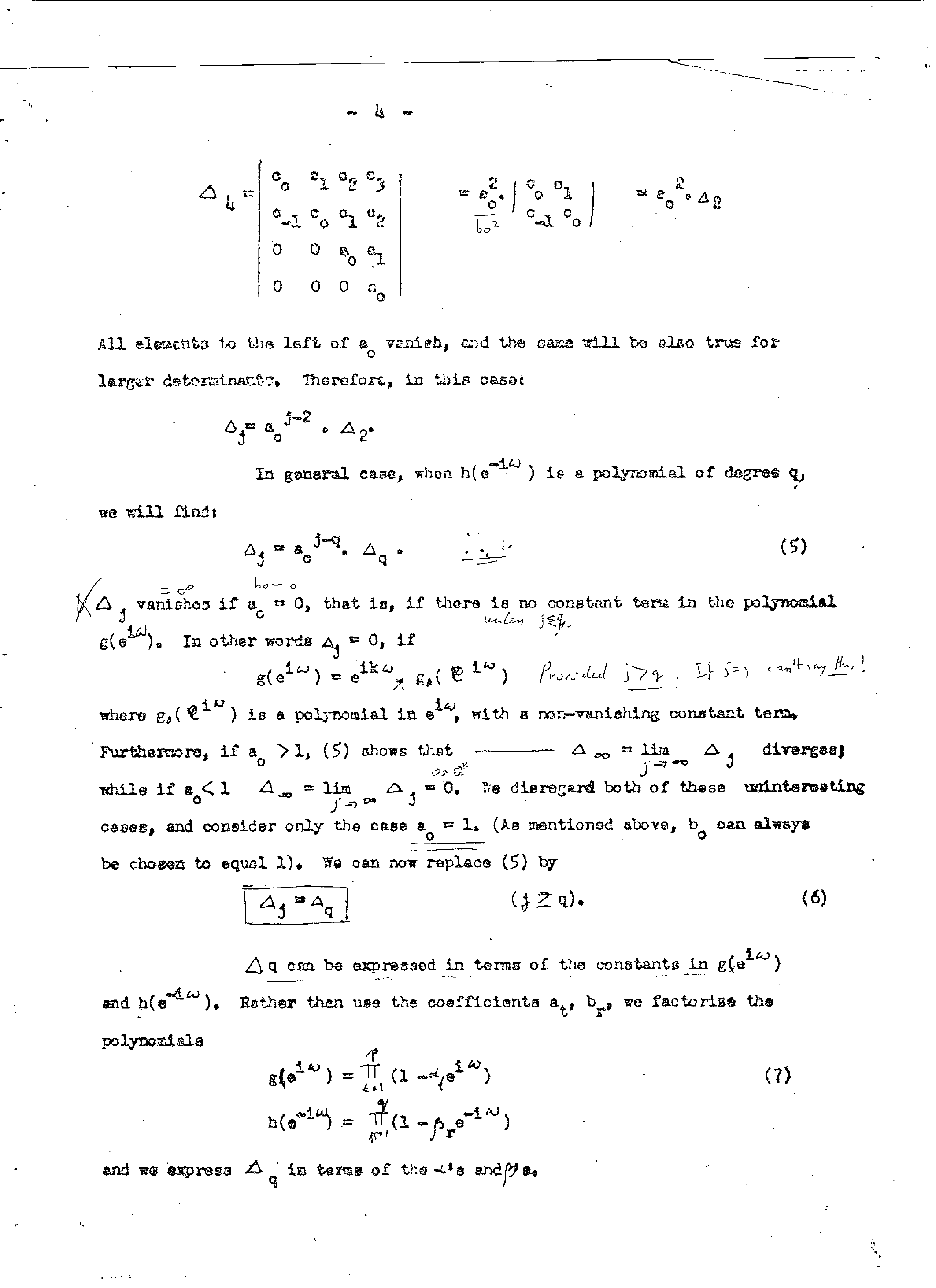}
\end{picture}
 \end{figure}
\newpage

\begin{figure}[hbt]
\begin{picture}(320,470) (14,200)
\setlength{\unitlength}{1.0pt}
 \includegraphics[height=25cm]{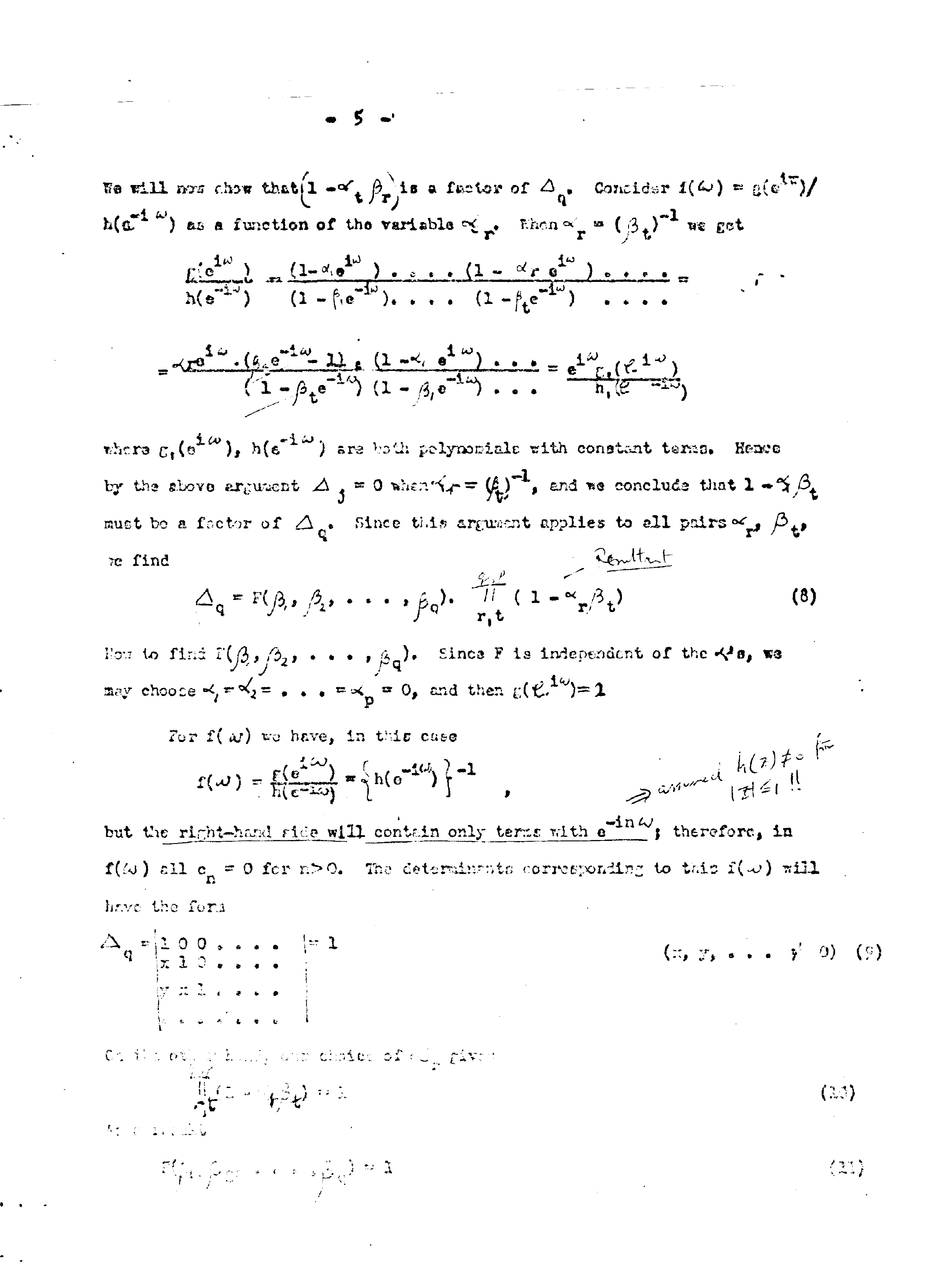}
\end{picture}
 \end{figure}
\newpage

\begin{figure}[hbt]
\begin{picture}(320,470) (14,200)
\setlength{\unitlength}{1.0pt}
 \includegraphics[height=25cm]{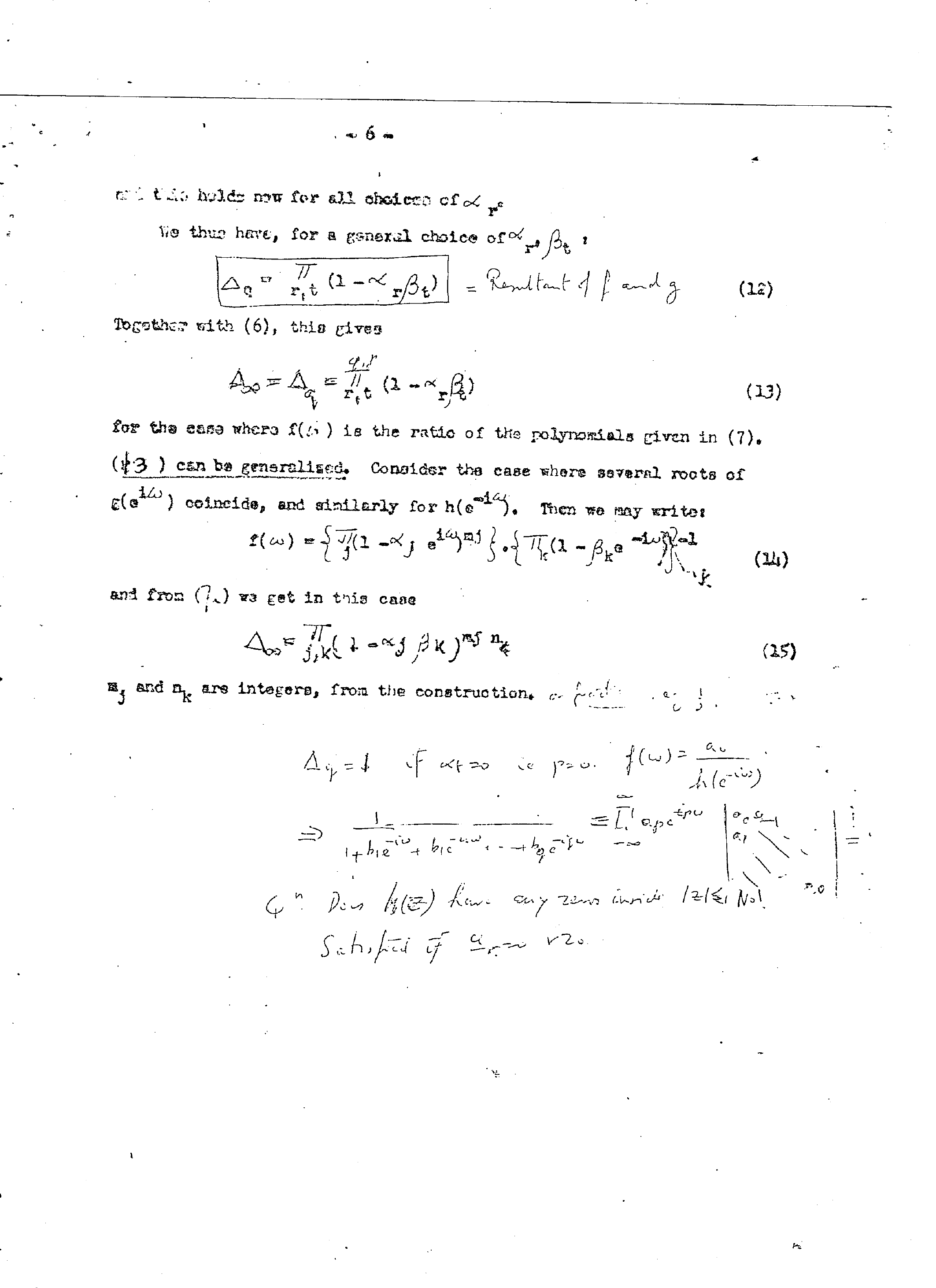}
\end{picture}
 \end{figure}
\newpage

\begin{figure}[hbt]
\begin{picture}(320,470) (14,200)
\setlength{\unitlength}{1.0pt}
 \includegraphics[height=25cm]{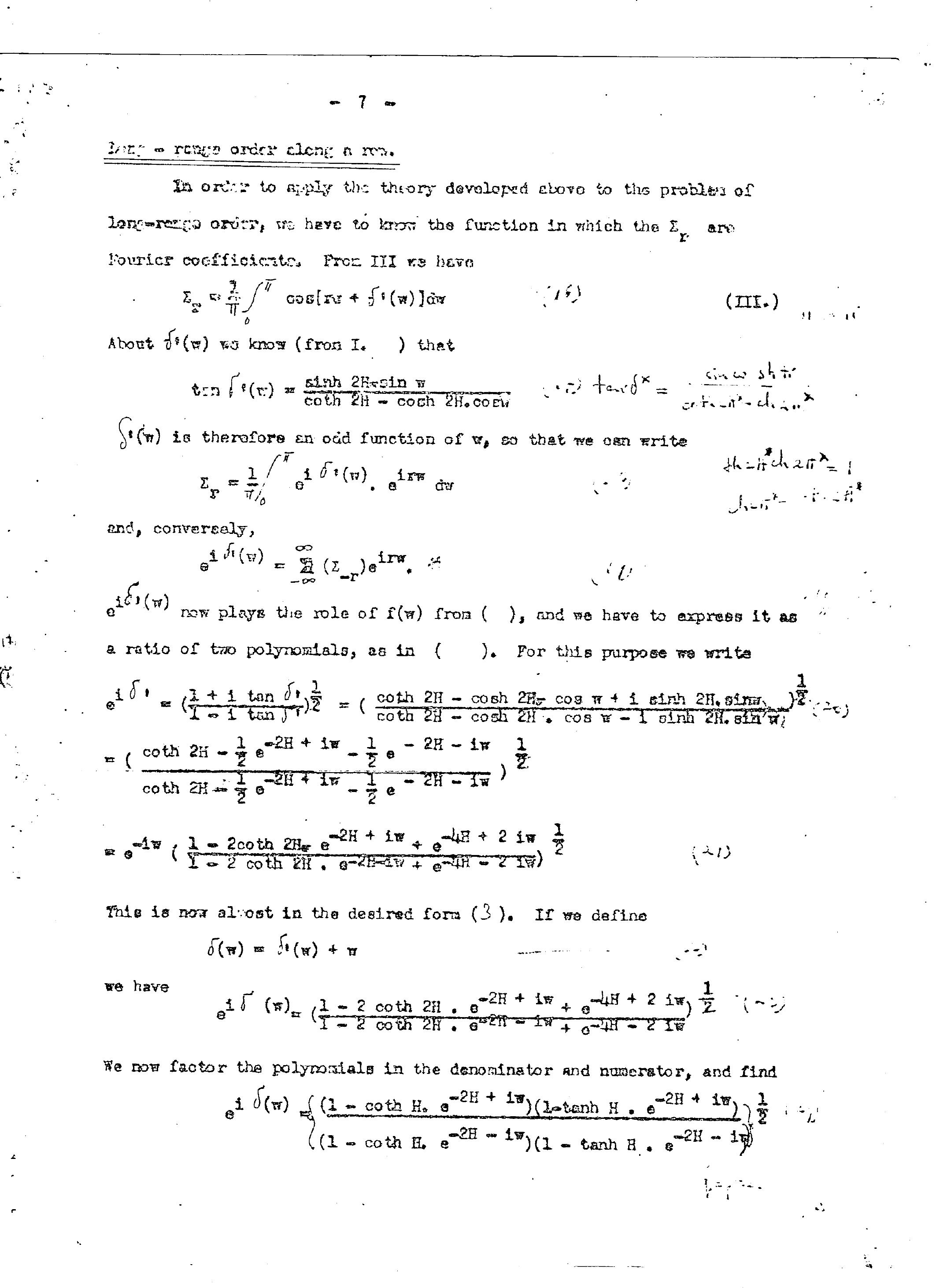}
\end{picture}
 \end{figure}
\newpage

\begin{figure}[hbt]
\begin{picture}(320,470) (14,200)
\setlength{\unitlength}{1.0pt}
 \includegraphics[height=25cm]{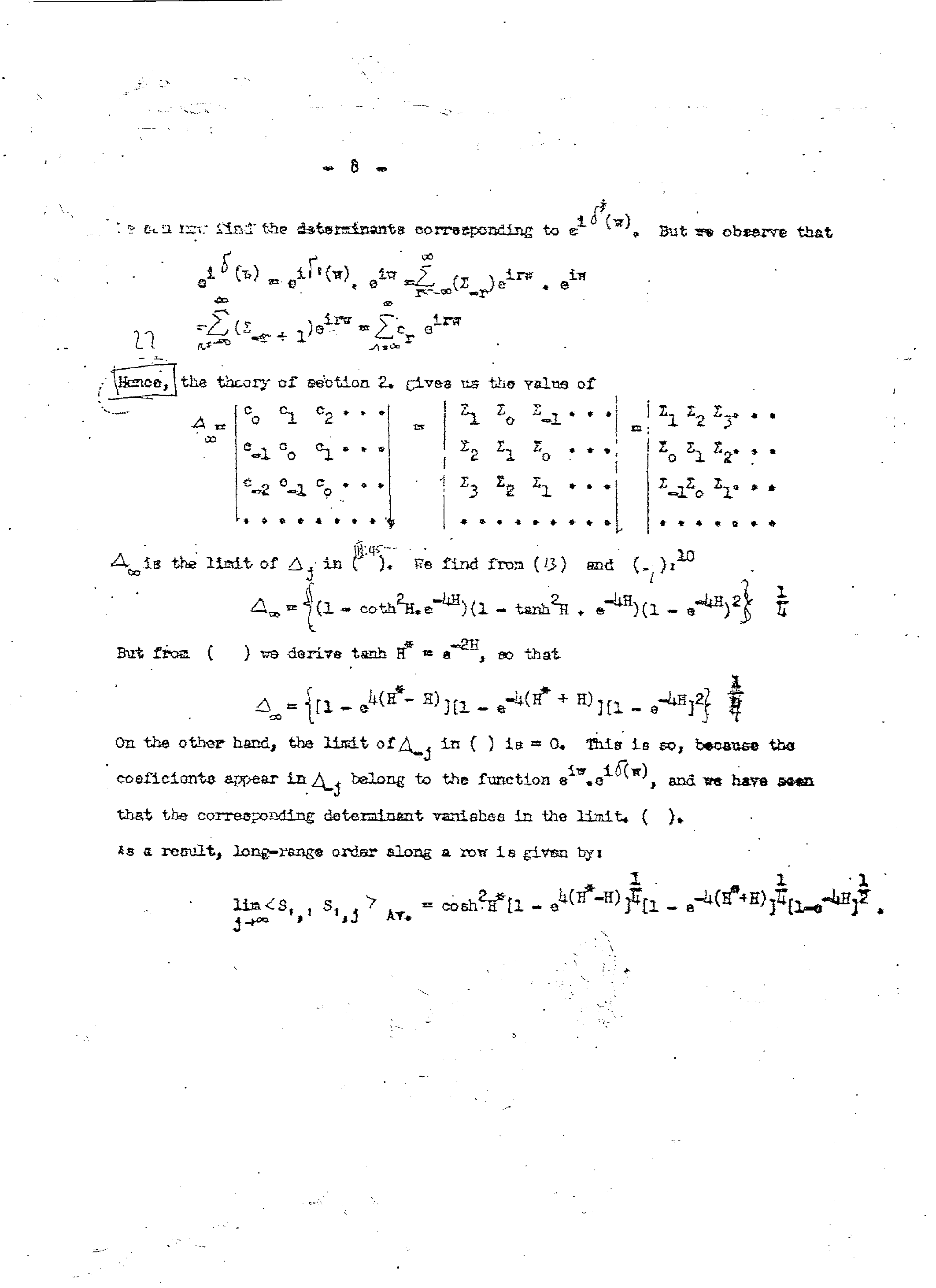}
\end{picture}
 \end{figure}
\newpage



\end{document}